# AoI-Energy-Spectrum Optimization in Post-Disaster Powered Communication Intelligent Network via Hierarchical Heterogeneous Graph Neural Network

**Hanjian Liu, Jinsong Gui, Xiaoheng Deng**

*Abstract*—**This paper proposes a post-disaster powered communication intelligent network (PDPCIN) designed to address communication disruptions caused by ground base station (GBS) failures in post-disaster areas. PDPCIN utilizes unmanned aerial vehicles (UAVs) to provide wireless data collection (WDC) and wireless energy transmission (WET) to affected areas, while leveraging low earth orbit satellites (LEO SATs) to relay UAV-collected data to the nearest operational GBS. To ensure fundamental post-disaster communication services and simultaneously optimize age of information (AoI), energy efficiency, and spectrum utilization, this study introduces an intelligent synchronization UAV (IS-UAV) architecture, an AoI-based four-threshold updating (AFTU) mechanism, and a dynamic multi-LEO access (DMLA) strategy. However, three major challenges persist: time-varying task-resource imbalances, complex topologies arising from multi-device scheduling, and nonlinear coupling in multi-dimensional metric optimization, which collectively render system optimization NP-hard. To tackle these issues, this paper presents a hierarchical heterogeneous graph neural networks (HHGNN) framework. The framework models heterogeneous devices and their communication relationships as a hierarchical heterogeneous graph, incorporating our defined graph sensing, exchange, and mask layers to manage input processing, feature propagation, and output generation within the graph network. Additionally, we propose a single-LEO SAT demand density optimization (S-LSDO) algorithm to determine the optimal number of LEO SATs required. Finally, we evaluate the proposed schemes against state-of-the-art benchmarks to demonstrate its superior performance in jointly optimizing AoI, energy efficiency, and spectrum utilization. Based on this analysis, we derive mathematical expressions for the expected values of AoI and the proportion of stagnant AoI.**

## I. INTRODUCTION

For an extended period, numerous areas have been affected by various natural disasters, resulting in substantial economic losses and posing significant threats to human safety. Ensuring reliable communication services in these disaster-stricken areas is essential for facilitating rescue operations, intelligence gathering, and post-disaster recovery [1]. In the aftermath of events such as earthquakes or floods, terrestrial communication infrastructure—including ground base stations (GBSs)—is often partially or entirely damaged. Due to the area-specific coverage characteristics of cellular networks, ground terminals (GTs) located in different zones are unable to establish direct connections with surviving GBSs, thereby disrupting communication in terrestrial networks following a disaster [2].

In this context, the space-air-ground integrated network (SAGIN) offers a promising solution to address these challenges. According to the International Telecommunication Union's (ITU) "Framework and overall objectives of the future development of IMT for 2030 and beyond," the establishment of a fully covered and intelligently coordinated SAGIN has emerged as one of the key objectives in the evolution of 6G standards [3]. Furthermore, the integration of unmanned aerial vehicles (UAVs) and low earth orbit satellites (LEO SATs) into network architectures has significantly accelerated the maturation of SAGIN technologies.

In ground-to-air (G2A) scenarios, UAVs—leveraging their high mobility and rapid deployment capabilities—are typically utilized as flying base stations or aerial relays to supplement or back up existing infrastructure, thereby extending communication coverage to GTs [4]. Consequently, UAVs can provide flexible communication services and deliver critical wireless energy to resource-constrained post-disaster environments. Within UAV-assisted wireless powered communication networks (WPCNs), recent studies have explored various aspects: [5] investigated federated learning and offloadable mobile edge computing tasks; [6] examined dynamic GT type updates; and [7] focused on minimizing the long-term average age of information (AoI). However, several challenges remain in post-disaster communication: due to the lack of prior knowledge regarding GT locations, UAVs must perform search operations to locate GTs; multi-UAV

Corresponding author：Jinsong Gui, Xiaoheng Deng

Hanjian Liu, Jinsong Gui, and Xiaoheng Deng are with School of Electronic Information, Central South University, Changsha, 410075 China (e-mail: {224712148, jsgui2010, dxh}@csu.edu.cn).

coordination is essential since individual UAV base stations possess limited sensing capabilities and can only detect nearby service demands; and frequent control signal updates are necessary due to time-varying network topologies and channel conditions caused by UAV movement and GT type changes [8], [9], [10].

In air-to-space (A2S) scenarios, LEO SATs offer distinct advantages over geostationary or medium earth orbit SATs, including reduced propagation loss and transmission delay. Their wide-area coverage capability enables them to relay data collected by UAVs to nearby ground base stations for further processing [11]. Studies [12], [13], [14] have respectively analyzed LEO SAT energy consumption, link outage probability, and data collection strategies. Nevertheless, integrating LEO SATs into post-disaster communication networks presents several challenges, including scheduling among multiple UAVs and SATs, bandwidth allocation across multiple SATs, and outage probabilities control [15], [16].

To summarize, three core challenges persist in optimizing post-disaster communication networks: 1) Imbalance between time-varying transmission demands and limited resources: The demand from GTs and UAVs for data transmission and resource supply fluctuates over time, necessitating efficient utilization of constrained resources to meet evolving service requirements. 2) Spatiotemporal conflicts between dynamic service scheduling and resource allocation: The service scheduling topology within the SAGIN architecture is complex, requiring rational and adaptive adjustments to resource distribution. 3) Nonlinear coupling among multi-dimensional performance metric optimizations: Key performance indicators such as AoI, energy efficiency, and spectrum utilization are interdependent but not linearly correlated, necessitating a balanced approach to achieve global optimization.

This paper proposes the integration of UAVs and LEO SATs into a post-disaster communication framework, forming the post-disaster powered communication intelligent network (PDPCIN). Given the intricate service and scheduling relationships among GTs, UAVs, and LEO SATs, conventional deep reinforcement learning (DRL) methods struggle to solve the problem effectively. In contrast, graph neural networks (GNNs) demonstrate superior capability in capturing the underlying structure of irregular data and modeling relationships among entities [17]. Moreover, due to the tight coupling between G2A and A2S scenarios, traditional GNN approaches struggle to achieve optimal network design. To overcome this limitation, we propose a novel hierarchical heterogeneous graph neural network (HHGNN) framework that sequentially addresses communication service issues in both G2A and A2S scenarios. Our main contributions are summarized as follows:

*1) Pragmatic System Network PDPCIN for Multi-Objective Optimization:* Section III models the post-disaster communication network and forms the SAGIN-based PDPCIN framework. Aiming at joint optimization of AoI, energy efficiency, and spectrum utilization, three key components are proposed: *a*) AoI-based four-threshold updating (AFTU) mechanism enables each GT to dynamically switch between data-transmitting mode and energy-harvesting mode in each time slot, taking into account AoI, communication system's non-stationarity, and type switching cost. *b*) Intelligent synchronization UAV (IS-UAV) architecture supports concurrent wireless energy transmission (WET) and wireless data communication (WDC) by autonomous WET decision-making, its collection-charge synchronization mode fully exploits the parallel capabilities of UAVs, enabling fine-grained resource allocation. To our best knowledge, this is the first attempt to study the WDC and WET simultaneous control and optimization in SAGIN architectures integrating WPCNs. *c*) Dynamic multi-LEO access (DMLA) strategy coordinates joint scheduling across multi-UAV and multi-LEO systems while accounting for A2S communication outage probability, thereby enhancing the robustness of post-disaster communication networks.

*2) Complex Communication Network Solution via HHGNN Framework:* Section IV formulates the global problem to achieve AoI-energy-spectrum collaborative optimization by jointly optimizing UAV's 3D continuous trajectories, WET decisions, UAV-LEO scheduling, UAV transmit power control, and SAT's subchannel allocation, under various constraints. However, solving this mixed-integer nonlinear programming (MINLP) problem independently is highly challenging due to its non-convex nature and combinatorial complexity. To address this issue, we propose the HHGNN architecture, which effectively decouples the global problem $\mathbb{GP}$ into two layered subproblems $\mathbb{L}1$ and $\mathbb{L}2$ corresponding to G2A and A2S scenarios. We further analyze the equivalence of decoupling problems and formulate MDPs on $\mathbb{L}1$ and $\mathbb{L}2$.

*3) G3M and S-LSDO algorithm:* Section V define graph sensing layer (GSL) to aggregate concatenated features with fluctuating dimensions, graph exchange layer (GEL) to alleviate high-overhead feature transmission, and graph mask layer (GML) to smooth and mask unavailable actions, respectively, which collectively constitute GSL-GEL-GML model (G3M). Additionally, single-LEO SAT demand density optimization (S-LSDO) algorithm is proposed to explore the deeper relation between SAT demand density of single-LEO and AoI/spectrum utilization performance. The algorithms' time complexity is calculated to verify their feasibility.

*4) Extensive Simulation Results and Analytical Derivations:* In Section VI, simulation results validate the superior performance of the proposed schemes, both in G2A and integrated G2A-space scenarios, compared to state-of-the-art benchmarks in terms of AoI, energy efficiency, and spectrum utilization. Moreover, iterative experiments are conducted to investigate the impact of varying stagnation-age of information (S-AoI) proportions and the number of subchannels on the required SAT density of single-LEO, where S-AoI proportion denotes the fraction of total AoI attributed to UAV waiting for SAT services. Based on these experimental findings, we further derive the analytical expressions for expected values of G2A-AoI, A2S-AoI, and S-AoI proportion, respectively.

The remainder of this paper is structured as follows: Section II reviews the related works. Then, we describe the system model in Section III, followed by HHGNN's problem and MDP formulation in Section IV. The design of neural networks and algorithms in HHGNN are elaborated in Section V, and the simulation results are analyzed while the expected values are derived in Section VI. Finally, we conclude this paper in Section VII.

## II. RELATED WORK

We begin with a review of existing research related to the UAV-aided WPCNs in G2A communication scenario. In [18], UAVs were divided into dedicated teams responsible for data collection and energy transmission, and a joint trajectory optimization approach was proposed. To enhance the number of covered devices, time efficiency, and energy utilization while minimizing flight distance, [19] formulated a joint UAV power and 3D trajectory optimization problem. With a focus on the charging process, [20] introduced a V-shaped WET scheme aimed at maximizing the energy harvested by GTs, as well as an inverted trapezoidal WET scheme designed to improve energy fairness among GTs. To address inefficiencies in WET caused by distance and environmental obstacles, [21] proposed a quality-of-experience-driven framework incorporating aerial intelligent reflective surfaces. In the context of interference management, [22] addressed co-channel and cross-link interference in multi-UAV WPCNs, whereas [5] focused on processing both federated learning tasks and offloadable mobile edge computing workloads. Furthermore, [6] proposed a multi-agent hierarchical DRL framework to support continuous trajectory planning and dynamic WDC/WET decision-making. For long-term average AoI minimization, [7] introduced a hybrid time division multiple access (TDMA) and non-orthogonal multiple access (NOMA) protocol combined with a clustering-based dynamic shortest path algorithm.

The aforementioned studies primarily rely on linear energy harvesting models, under which GTs are assumed to harvest non-zero energy regardless of the intensity or aggregation level of the received signals. Additionally, these works typically divide the mission period into sequential WET and WDC phases, which simplifies system design but fails to account for the heterogeneous requirements of GTs regarding energy acquisition and data transmission. Importantly, since GTs utilize UAV-supplied energy for uplink transmissions, WET and WDC exhibit interdependent trade-offs that significantly impact overall system performance. Although [6] addresses the limitation in [18] by enabling adaptive WET and WDC decisions for each UAV, its time-slotted decision-making mechanism underutilizes the concurrent capabilities of UAVs. While [19] and [5] equip UAVs with multiple orthogonal isotropic antennas and radio frequency transmitters to enable simultaneous WDC and WET, they lack autonomous WET control mechanisms responsive to dynamic environmental conditions, leading to considerable energy wastage due to continuous transmissions.

We next examine SAGIN works integrating LEO SATs. Given the characteristics of seamless coverage and global broadband access offered by SATs, SAGINs integrating UAVs and SATs have attracted significant research interest. In [23], a hierarchical bandwidth allocation scheme was proposed to support high-quality multicast services within SAGIN-based social communities. Meanwhile, [24] minimized data collection completion time through joint optimization of UAV trajectory, Internet of Remote Things (IoRT) device association, and data caching strategies. [25] investigated optimal task offloading strategies and resource allocation for mobile edge computing within SAGIN environments via considering task dependencies. [26] introduced an end-to-end network slicing architecture for control- and user-plane separated SAGINs. To minimize time-averaged network costs, [12] proposed a perception-aided online DRL approach. [27] explored cognitive radio-enabled reconfigurable intelligent surface-assisted NOMA-based SAGINs. By jointly considering UAV channel fading, energy consumption, and energy harvesting dynamics, [13] developed an integrated analytical model for SAGIN transmission performance. [14] designed an ISS-proximal policy optimization algorithm for resource allocation in sink-UAV-LEO data collection architectures.

Despite achieving notable advancements, the above studies overlook the challenges associated with multi-UAV to multi-LEO SAT scheduling and the resulting complex topologies inherent in multi-hop communication scenarios. Especially, dynamic and heterogeneous graph-based topology modeling for SAGIN remains underexplored. As summarized in Table I, SAGIN architectures that integrate WPCNs have been rarely investigated in existing literature. Moreover, most UAV path planning schemes adopt discrete action spaces or neglect altitude optimization for simplicity. In contrast, continuous 3D trajectory optimization enables finer-grained control over WPCN operations but introduces increased topological complexity. Consequently, research that simultaneously addresses dynamic heterogeneous network topology modeling, optimization of specific communication bottlenecks, and coordinated multi-objective enhancement within WPCN-SAGIN resource allocation remains limited and presents significant technical challenges—motivating the current study.

TABLE I: COMPARISON BETWEEN OUR WORK AND EXISTING WORKS

| Novelty | [18] 2022 | [19] 2023 | [20] 2023 | [21] 2024 | [22] 2024 | [5] 2024 | [6] 2024 | [7] 2025 | [23] 2023 | [24] 2024 | [25] 2024 | [26] 2024 | [12] 2024 | [27] 2025 | [13] 2025 | [14] 2025 | Our work |
|---|---|---|---|---|---|---|---|---|---|---|---|---|---|---|---|---|---|
| WDC decision | ✓ | ✓ |  | ✓ | ✓ | ✓ | ✓ | ✓ | ✓ | ✓ | ✓ | ✓ | ✓ | ✓ | ✓ | ✓ | ✓ |
| WET decision | ✓ |  | ✓ | ✓ | ✓ |  | ✓ |  |  |  |  |  |  |  |  |  | ✓ |
| WDC/WET simultaneity |  | ✓ |  |  |  | ✓ |  |  |  |  |  |  |  |  |  |  | ✓ |
| Non-linear energy harvester |  |  |  |  |  |  | ✓ |  |  |  |  |  |  |  |  |  | ✓ |
| GT type updating |  |  |  |  | ✓ |  | ✓ | ✓ |  |  |  |  |  |  |  |  | ✓ |
| UAV 3D trajectory | ✓ | ✓ |  |  |  |  |  |  |  |  | ✓ |  |  |  | ✓ |  | ✓ |
| UAV energy harvest |  |  | ✓ |  |  |  |  |  |  |  |  |  | ✓ |  | ✓ | ✓ | ✓ |
| SAT |  |  |  |  |  |  |  |  | ✓ | ✓ | ✓ | ✓ | ✓ | ✓ | ✓ | ✓ | ✓ |
| Multiple LEO |  |  |  |  |  |  |  |  |  |  |  |  |  |  |  |  | ✓ |
| SAT coverage |  |  |  |  |  |  |  |  |  | ✓ | ✓ |  |  |  | ✓ |  | ✓ |
| Outage probability |  |  |  |  |  |  |  |  |  |  |  |  |  |  | ✓ |  | ✓ |
| Network topology |  |  |  |  |  |  |  |  |  |  | ✓ |  |  |  |  |  | ✓ |
| AoI minimization | ✓ |  |  |  |  |  |  | ✓ |  |  |  |  |  |  |  |  | ✓ |
| Energy efficiency maximization |  | ✓ | ✓ |  |  | ✓ |  |  |  |  | ✓ | ✓ | ✓ | ✓ |  | ✓ | ✓ |
| Spectrum utilization maximization |  |  |  |  |  |  |  |  | ✓ |  |  |  |  |  |  | ✓ | ✓ |

## III. SYSTEM MODEL

In this section, we first introduce the PDPCIN architecture, and then model its G2A and A2S scenarios, respectively.

### *A. PDPCIN Architecture*

As illustrated in Fig. 1, the PDPCIN system incorporates both G2A and A2S scenarios. In the G2A scenario, multiple UAVs are deployed to post-disaster areas to collaboratively collect data from GTs. The sets of GTs and UAVs are denoted by $\mathcal{V}^G = \{1, \cdots, n, \cdots, N\}$ and $\mathcal{V}^U = \{1, \cdots, m, \cdots, M\}$, respectively. Considering the resource limitations commonly encountered in post-disaster environments, UAVs perform downlink WET to supply GTs with necessary energy. Meanwhile, GTs can operate in either T-GT mode (data transmission) or E-GT mode (energy harvesting).

Each GT is equipped with a single antenna and a rechargeable battery, enabling it to harvest energy from UAVs or transmit data within each time slot. To mitigate co-channel interference between downlink WET and uplink WDC, each UAV employs separate antennas operating on orthogonal frequency bands [6], [18]. This configuration allows simultaneous and independent execution of energy transfer and data collection, supporting autonomous WET decisions.

The proposed intelligent synchronization UAV (IS-UAV) architecture offers several advantages over conventional approaches such as static resource partitioning used in team division UAVs (TD-UAVs) [18] or serial processing in dynamic conversion UAVs (DC-UAVs) [6]. These benefits include: leveraging UAVs' parallel processing capabilities to enhance resource utilization; enabling fine-grained resource allocation for improved system adaptability; and preventing communication outages caused by energy depletion, thereby enhancing overall system robustness. Comparative experimental evaluations are presented in Section VI.

Each UAV is equipped with a replaceable battery that can be swapped once depleted, allowing the resumption of data collection tasks. Notably, due to the lack of prior knowledge regarding GT locations, limited UAV battery capacity, and inherent constraints in sensing range and coverage, trajectory planning plays a critical role in optimizing both WDC and WET operations. Additionally, UAVs are integrated with energy harvesting modules to capture renewable energy sources, such as solar power. To efficiently manage limited battery resources, ensure uninterrupted data transmission, and support subsequent neural network-based decision-making based on real-time energy status, the replaceable battery is exclusively allocated for non-communication functions (e.g., UAV mobility and energy transmission), while harvested energy is reserved for communication operations—specifically, for transmitting data to LEO SATs.

In the A2S scenario, an ultra-dense LEO SAT constellation is considered, comprising multiple LEO SATs covering the target area. Each LEO SAT periodically passes over the area, ensuring continuous coverage. Consequently, UAVs can select one LEO SAT for access in each time slot from among the available observable LEO SATs. These LEO SATs relay the collected data from UAVs to unaffected GBSs for further processing. UAVs are capable of simultaneously receiving data from GTs and transmitting data to LEO SATs using a store-and-forward mechanism, whereby data received from GTs in one time slot is forwarded to LEO SATs in the subsequent time slot. To improve spectral efficiency and system capacity, UAVs utilize NOMA when accessing LEO SATs. It is assumed that G2A communications operate in the *C*-band, while A2S communications use the *Ka*-band, ensuring no mutual interference [28].

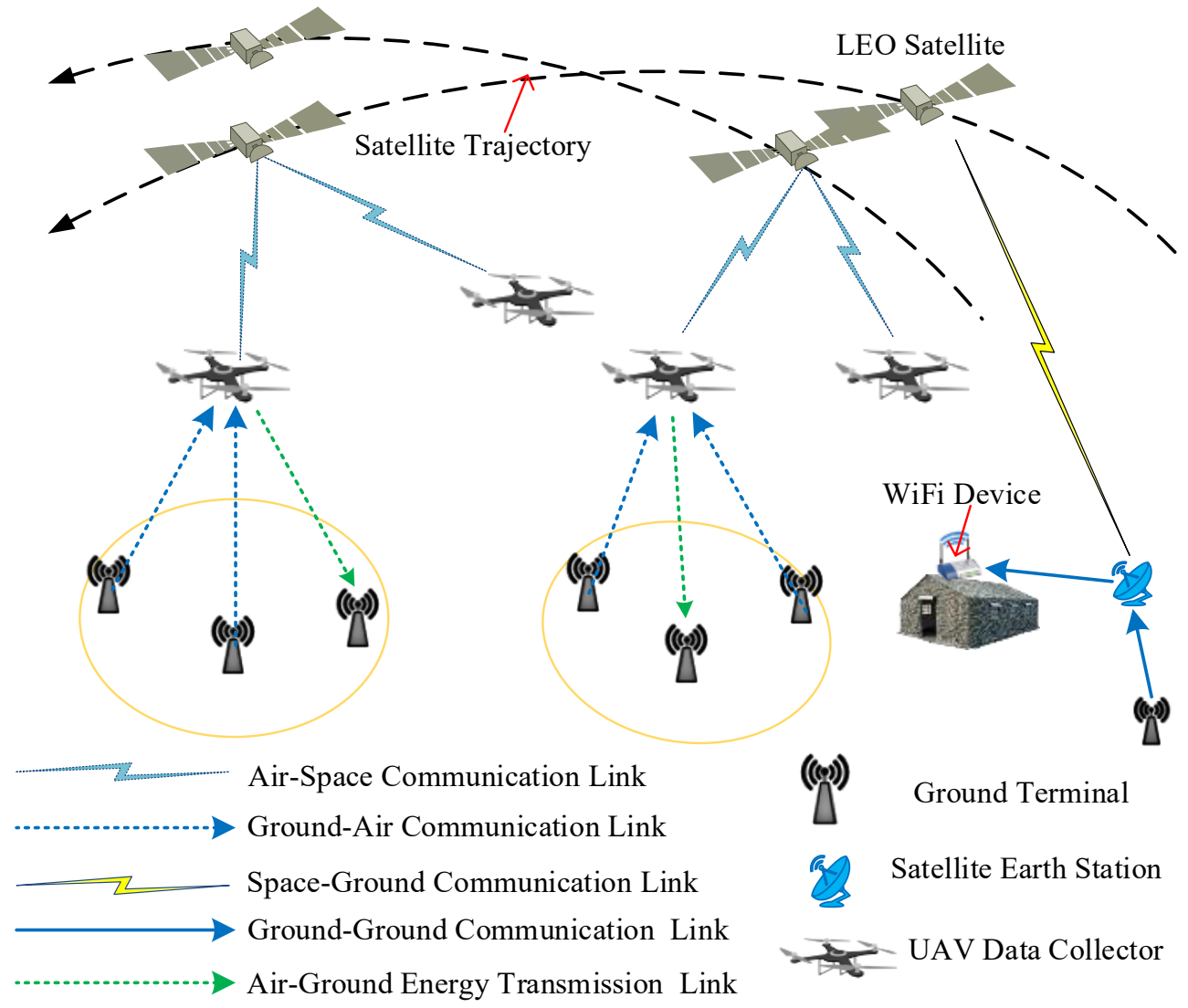

Fig. 1. Post-disaster powered communication intelligent network.

### *B. G2A Transmission Scenario Description*

#### *1) Sensing and Coverage Model*

We assume the position of GT-$n$ is denoted by $\boldsymbol{q}_n^G(t) = (x_n^G(t), y_n^G(t), z_n^G(t)) \in \mathbb{R}^3$. In slot $t$, the position of UAV-$m$ is denoted by $\boldsymbol{q}_m^U(t) = (x_m^U(t), y_m^U(t), z_m^U(t)) \in \mathbb{R}^3$. Moreover, we assume that all GTs are remain on the level ground, i.e., $z_n^G = 0$. The flight altitude of UAVs is dynamically adjusted. GTs transmit service requests and location coordinates via a control and non-payload communications (CNPC) link. Due to signal attenuation, UAVs can serve GTs only within a limited range. Similarly, each UAV can sense other UAVs within a certain range [29]. The maximum ranges for UAVs to detect GTs and other UAVs are denoted by $O_S^G$ and $O_S^U$, respectively. In slot $t$, UAV-$m$ can detect GT-$n$ only if $d_{n,m} \leq O_S^G$, where $d_{n,m}(t) = \|\boldsymbol{q}_m^U(t) - \boldsymbol{q}_n^G(t)\|$ is their mutual distance. UAV-$m$ and UAV-$i$ can detect each other only if their mutual distance $d'_{m,i}(t) = \|\boldsymbol{q}_m^U(t) - \boldsymbol{q}_i^U(t)\|$ satisfies $d'_{m,i} \leq O_S^U$. Moreover, UAV-m can collect data from GT-n or GT-n can harvest energy from UAV-m only if $d_{n,m}(t) \leq O_C^U$, where $O_C^U$ is UAVs' coverage range satisfying $O_C^U < O_S^G$. Consequently, each UAV has only partial system observation. Since UAVs hover over the area during each time interval, their observations change over time.

To describe the connectivity among GTs and UAVs at each slot $t$, we define three matrices $\mathbf{G}(t) \in \{0,1\}^{N\times M}$, $\mathbf{U}(t) \in \{0,1\}^{M\times M}$, and $\mathbf{C}(t) \in \{0,1\}^{N\times M}$. In slot $t$, $(n, m)$-th entry of $\mathbf{G}(t)$, denoted by $g_{n,m}(t)$, is 1 if GT-n is detected by UAV-$m$; similarly, $u_{i,m}(t) = 1$ if UAV-$i$ is detected by UAV-$m$; $c_{n,m}(t) = 1$ if GT-$n$ is within UAV-$m$'s coverage range and can be served by UAV-$m$; all are 0 otherwise.

*2) AoI-based Four-Threshold Updating*

UAVs evaluate information freshness using the AoI metric for each GT. In PDPCIN system, GTs' data packet generation follows the Poisson distribution, we refine the calculation of AoI to the packet level, and GT-$n$'s AoI $a_n^G(t)$ is computed as

$$a_n^G(t) = \sum_{i=1}^{I_n^G(t)} a_{n,i}^G(t), \quad (1)$$

$$a_{n,i}^G(t) = \begin{cases} 0, & \text{if received,} \\ a_{n,i}^G(t-1)+1, & \text{otherwise,} \end{cases} \quad (2)$$

where $a_{n,i}^G(t)$ is the AoI of GT-n's $i$-th packet at the end of slot $t$, and $I_n^G$ is the total number of packets generated by GT-$n$ up to the end of slot $t$. The average AoI of all GTs in slot $t$ is represented as $A^G(t) = \frac{\sum_{n=1}^{N} a_n^G(t)}{N}$. It is worth noting that the above GTs' AoI calculation method is also applicable to UAVs, with $a_m^U(t)$ denotes UAV-$m$'s AoI and $A^U(t)$ denotes average AoI of all UAVs in slot $t$ respectively.

To allow GTs with higher AoI to transmit data more quickly and reserve energy for those with less urgent tasks, we propose AFTU to update GT types. Denote $\epsilon_n(t)$ as GT-$n$'s type, where 0 represents a T-GT and 1 an E-GT. According to AGTU rule, GT-$n$ updates its type at the end of slot $t$ by (3). Let $B_n^G(t)$ denote GT-$n$'s battery energy level at the end of slot $t$, with threshold $B_E$ (E-GT) and $B_I$ (T-GT) satisfying $P^G\tau < B_E < B_I$, where $P^G$ and $\tau$ are T-GT's fixed transmit power and time slot duration. Coefficient $\xi \in [0,1]$ is the weight of GT type updating activity, with a lower value indicating increased frequent GT type updating to adapt to a non-stationary environment. From (3), GT type updating follows four thresholds: 1) power outside $B_E$ or $B_T$ triggers an update; 2) if not, update if AoI exceeds $(1 \pm \xi)A^G(t)$. Otherwise, the remains unchanged. Notably, global $A^G(t)$ is typically replaced by local $\hat{A}^G(t)$ calculated by GTs based on all GTs' AoI within their perceptible range, as global values are usually inaccessible.

It is worth mentioning that unlike the widely-used single-threshold method (causing frequent type switching), e.g., [22], or the newly proposed double-threshold based GT type updating approach (reducing updates via fixed thresholds) [6], our dynamic $\xi$ scaling in (3) ensures efficient AoI control in non-stationary environments while minimizing unnecessary type switches in stable conditions, improving flexibility and reliability.

$$\epsilon_n(t) = \begin{cases} 0, & if\ B_n^G(t) \geq B_T\ or\ \left(B_E < B_n^G(t) < B_T\ and\ a_n^G(t) \geq (1+\xi)A^G(t)\right), \\ 1, & if\ B_n^G(t) \leq B_E\ or\ \left(B_E < B_n^G(t) < B_T\ and\ a_n^G(t) \leq (1-\xi)A^G(t)\right), \\ \epsilon_n(t-1), & if\ B_E < B_n^G(t) < B_T\ and\ (1-\xi)A^G(t) < a_n^G(t) < (1+\xi)A^G(t). \end{cases} \quad (3)$$

*3) G2A Communication Model*

As all UAVs share the spectrum, overlapping coverage causes inter-UAV interference, thus degrading link quality. Therefore, each GT always select the nearest UAV covering it. Define a scheduling matrix $\mathbf{S}(t)$ with entry $s_{n,m}(t) = 1$ if GT-$n$ is scheduled to UAV-$m$ in slot $t$, else 0. Each GT accesses at most one UAV per slot, satisfying: $\sum_{m=1}^{M} s_{n,m}(t) \leq 1$. A frequency division multiple access (FDMA) scheme is adopted in G2A scenario, where the total GT-UAV spectrum is divided into $Y^U$ equal subchannels of width $W^U$, and each GT can be assigned to at most one subchannel. The uplink signal-to-interference-and-noise ratio (SINR) from GT-$n$ to UAV-$m$ is computed by

$$\gamma_{n,m}(t) = \frac{s_{n,m}(t)h_{n,m}(t)P^G}{\sum_{i\in\mathcal{V}^G, i\neq n}\sum_{j\in\mathcal{V}^U} c_{i,m}(t)s_{i,j}(t)h_{i,m}(t)P^G + \mathbb{N}_0 W^U}, \quad (4)$$

where $\mathbb{N}_0$ is the power spectral density of the additive white Gaussian noise (AWGN), $P^G$ is GTs' fix transmit power, and $h_{i,m}(t)$ is the channel gain between GT-$n$ and UAV-$m$ in slot $t$, calculated via the line-of-sight (LoS)/non-line-of-sight (NLoS) channel model in [30]. The data volume transmitted by GT-$n$ and collected by UAV-$m$ in slot $t$ are obtained as

$$D_n^G(t) = \sum_{m=1}^{M} W^U \log\left(1+\gamma_{n,m}(t)\right)\tau, \quad (5)$$

$$D_m^U(t) = \sum_{n=1}^{N} W^U log\left(1+\gamma_{n,m}(t)\right)\tau. \quad (6)$$

Accordingly, the data volume collected by UAV-$m$ in slot $t$ denotes by $D_m^U = \sum_{n=1}^{N} s_{n,m} D_n^G(t)$. To address the situation where scheduled GTs outnumber available subchannels, we propose a AoI-fairness metric $\varrho_n(t)$ (with a higher value indicating higher priority) that jointly considers AoI and service fairness. It is calculated by $\varrho_n(t+1) = \frac{\varkappa a_n^G(t)}{\widehat{A^G}(t)} - \frac{(1-\varkappa)N\widehat{D}_n^G(t)}{\widehat{D}^G(t)}$, where $\widehat{D}_n^G(t) = \sum_{i=1}^{t} D_n^G(i)$ is total data amount transmitted by GT-$n$ up to slot $t$, $\widehat{D}^G(t) = \sum_{n=1}^{N} c_{n,m}(t)\widehat{D}_n^G(t)(1-\epsilon_n(t))$ is total transmitted data amount up to slot $t$ of GTs which request to access UAV-$m$ in slot $t+1$, and $\varkappa \in [0,1]$ is the AoI-fairness balance coefficient with lower values tending to fair scheduling and higher values tending to low-AoI scheduling. Furthermore, we employ long-term Jain's fairness index to quantify the fairness of UAV services, computed as $F(t) = \frac{\left(\sum_{n=1}^{N}\widehat{D}_n^G(t)\right)^2}{N\sum_{n=1}^{N}\left(\widehat{D}_n^G(t)\right)^2}$, where the $F(t) \in \left[\frac{1}{N}, 1\right]$, with higher values indicating greater fairness.

*4) G2A Energy Model*

First, we explore the GT energy harvest model. Let $Z_m(t) \in \{0,1\}$ denote UAV-$m$'s WET decision in slot $t$: it transmits energy with fixed power $P_E^U > 0$ only if $Z_m(t) = 1$. Each E-GT uses a non-linear energy harvester that converts received radio frequency (RF) power $p_e$ into stored direct current (DC) power $\overline{P}(p_e)$ via a non-linear function in [31]. E-GT-$n$'s harvested energy in slot $t$ is

$$E_n^{Gh}(t) = \epsilon_n(t)\bar{P}\left(\sum_{m=1}^{M} P_E^U Z_m(t)h_{n,m}(t)\right)\tau. \quad (7)$$

According to [32], UAVs with $Z_m(t) = 1$ must locate near E-GT-$n$ for non-zero energy harvesting. Denote $B_{max}^G$ as the GT battery capacity and $E_n^{Gc}(t) = \sum_{m=1}^{M} s_{n,m}(t)P^G\tau$ as T-GT-$n$'s energy consumption in slot $t$, GT-$n$'s remaining energy at the beginning of time slot $t+1$ is updated as

$$B_n^G(t+1) = min(B_n^G(t) + E_n^{Gh}(t) - E_n^{Gc}(t), B_{max}^G). \quad (8)$$

Due to $P^G\tau < B_E$ in AGTU to prevent GTs from transmitting data under insufficient energy, $B_n^G(t+1)$ will not be lower than 0. Next, we investigate the UAV energy consumption model. In G2A scenario, UAV energy consumption primarily arises from movement, WET, and WDC. UAV-$m$'s total energy consumption in slot $t$ is:

$$E_m^{Uc}(t) = (\sum_{n=1}^{N} s_{n,m}(t)P_C^U + P_{pro}\left(V_m^U(t)\right) + Z_m(t)P_E^U)\tau, \quad (9)$$

where $P_C^U$ is the fixed WDC consumption power at each UAV, and $P_{pro}\left(V_m^U(t)\right)$ is UAV-$m$'s propulsion power consumption in slot $t$, which is estimated according to the model in [30] and the corresponding velocity is estimated by $V_m^U(t) \triangleq \frac{1}{\tau}\|q_m^U(t+1)-q_m^U(t)\|$. Accordingly, UAV-$m$'s battery energy level at the beginning of slot $t+1$ is calculated as

$$B_m^U(t+1)=\max\left(B_m^U(t)-E_m^{Uc}(t),0\right). \tag{10}$$

It is assumed that every UAV starts with a full charge, such that $B_m^U(0)=B_{max}^U$, where $B_{max}^U$ denotes UAV battery capacity.

### C. A2S Transmission Scenario Description

We consider a set $\mathcal{K}=\{1,\cdots,k,\cdots,K\}$ of LEOs and sets $\mathcal{V}^k=\{1,\cdots,l^k,\cdots,L^S\}$ of LEO SATs in LEO-$k$, with $|\mathcal{V}^k|=L^S,\forall k\in\mathcal{K}$. Without loss of generality, assume all SATs move in uniform circular motion with period $\mathcal{T}$, and arc length between any adjacent LEO SATs are equal. SATs in each LEO serve the area of interest uninterruptedly in turns, with each UAV attempting to access only to the closest LEO SAT in each LEO [34]. The area of interest (e.g., 2 km radius) is much smaller than the SAT beam coverage range (e.g., 580 km radius) [35] and the arc length between adjacent LEO SATs in the same LEO (e.g., 1976 km). Thus, for all UAVs, the nearest LEO SAT per orbit is identical within a given time slot.

In DMLA system, the spectrum allocation mode, adopted to enhance spectrum utilization and system capacity, involves three key steps: 1) Dynamic SAT-UAV Connection: each UAV is strategically scheduled to the corresponding LEO and establishes connection with the LEO's closest SAT dynamically based on their specific communication requirements, ensuring adaptive resource utilization. 2) Subchannel Partitioning: The available frequency band is divided into smaller subchannels to enable flexible allocation and sharing. These subchannels are strategically assigned in different proportions to each LEO's SAT closest to the area of interest. 3) NOMA-based Subchannel Sharing: UAVs connected to the same LEO SAT employ NOMA technology to share the allocated subchannels at dynamic transmit power, maximizing spectrum reuse. Details of UAV-LEO scheduling, subchannel allocation, and UAV power control strategies are provided in Section IV and V.

*1) Single-LEO Coverage Model*

We employ single-LEO coverage model to derive the per-cycle SAT service duration for UAVs, with the multi-LEO case following similarly. The angle between the UAV-$m$ and SAT-$l^k$ can is obtained as [25]

$$\omega_A=\arccos\left(\frac{d_E+h_k}{\hat{d}_{m,l^k}}\sin\omega_C\right), \tag{11}$$

where $d_E$ is earth radius, $h_k$ is LEO-$k$'s height, $\hat{d}_{m,l^k}$ is the distance between UAV-$m$ and SAT-$l^k$, and $\omega_C$ is the angle for SAT-$l^k$'s coverage. Note that SATs cease service to UAVs when their elevation angle falls below the predefined minimum threshold $\omega_A$. $\omega_C$ is given by

$$\omega_C=\arccos\left(\frac{d_E}{d_E+h_k}\cos\omega_A\right)-\omega_A. \tag{12}$$

Since the movement range of UAVs is negligible compared to $d_E$ and $h_k$, the coverage time $T_{m,l^k}$ of UAV-$m$ by SAT-$l^k$ is calculated as

$$T_{m,l^k}=\frac{2(d_E+h_k)\omega_C}{V^S}, \tag{13}$$

where $V^S$ is LEO SATs' orbital velocity.

*2) Multi-LEO NOMA Model*

In wireless communication, outage probability, the likelihood that the instantaneous rate drops below a threshold, reflects link reliability under varying channel conditions. In NOMA systems, its importance is amplified by non-orthogonal resource sharing, which intensifies inter-user interference to elevate AoI. Thus, managing outage probability is vital for ensuring high-reliability communication [36]. Define a outage matrix $\mathcal{O}(t)$ of which the $(m,k,l)$-th entry $o_m^{l^k}(t)=0$ if $R_m^{l^k}(t)\geq\hat{R}_m^{l^k}$ and 1 otherwise, where $R_m^{l^k}(t)$ is the transmission rate from UAV-$m$ to SAT-$l^k$, and $\hat{R}_m^{l^k}$ is the minimum data rate required for SAT-$l^k$ to decode UAV-$m$'s signal [37]. Intuitively, coordinating the rates of all UAVs is crucial to reduce outages: excessively high rates intensify interference, disrupting other UAVs' decoding, while too low rates cause self-outages.

Denote $\mathcal{J}(t)=\{J^1(t),\cdots,J^k(t),\cdots,J^K(t)\}$ as the number of UAVs accessing to each LEO's SATs in slot $t$. To mitigate successive interference cancellation (SIC) process's interference effects, assume the channel gains of UAVs accessing the same SAT be sorted in descending order as $|\hbar_1^{l^k}(t)|^2\geq|\hbar_2^{l^k}(t)|^2\geq\cdots|\hbar_m^{l^k}(t)|^2\geq\cdots|\hbar_{J^k(t)}^{l^k}(t)|^2$.

During decoding of UAV-$m$'s message, signals from UAV-$i$ $(i<m)$ are canceled, while signals from UAV-$j$ $(j>m)$ are treated as noise. Define a scheduling matrix $\boldsymbol{S}(t)$ of which the $(m,k,l^k)$-th entry $s_m^{l^k}(t)=1$ if UAV-$m$ connects to SAT-$l^k$ in slot $t$ and 0 otherwise, satisfying $\sum_{k=1}^K\sum_{l^k\in\mathcal{V}^k}s_m^{l^k}(t)\leq 1$. More, assume there are $Y^S$ subchannels between UAVs and LEO SATs, where $\{\rho_1(t),\ldots,\rho_k(t),\ldots,\rho_K(t)\}$ denotes the subchannel ratios (with $\sum_{k\in\mathcal{K}}\rho_k\leq 1$) of each LEO's allocated subchannels to the total subchannels in slot $t$. Each UAV's total transmit power is evenly distributed across allocated subchannels. Considering outage probability, the SINR of UAV-$m$ connected to SAT-$l^k$ in each subchannel is calculated as:

$$\gamma_m^{l^k}=\frac{s_m^{l^k}(t)\hbar_m^{l^k}(t)P_m^U(t)}{\sum_{i=m+1}^{J^k(t)}\hbar_i^{l^k}(t)P_m^U(t)+\sum_{i=1}^{m-1}o_m^{l^k}(t)\hbar_i^{l^k}(t)P_m^U(t)+W^S\mathbb{N}_0}, \tag{14}$$

where $W^S$ is UAV-SAT subchannel bandwidth, $P_m^U(t)$ is UAV-$m$'s per-subchannel transmit power, and $\hbar_m^{l^k}(t)$ is the LoS/NLoS channel gain between UAV-$m$ and SAT-$l^k$ in slot $t$ calculated according to [38], [39], [40]. Accordingly, UAV-$m$'s transmission capacity and its actual throughput in slot $t$ can be denoted as:

$$\dot{D}_m^U(t)=Y^SW^S\sum_{k=1}^K\sum_{l^k\in\mathcal{V}^k}\rho_k(t)log_2\left(1+\gamma_m^{l^k}\right)\hat{\tau}_m(t), \tag{15}$$

$$\hat{D}_m^U(t)=\min\left(D_m^U(t),\dot{D}_m^U(t)\right), \tag{16}$$

where $\hat{\tau}_m(t)=\min\left(\frac{\hat{B}_m^U(t)+E_m^{Uh}(t)}{P_m^U(t)},\tau\right)$ is the actual duration of data transmission, $\hat{B}_m^U(t)$ is energy storage of UAV-$m$'s energy harvesting board at the beginning of slot $t$, and $E_m^{Uh}(t)$ is UAV-m's harvested energy in slot $t$. UAV-$m$'s total transmission energy consumption in slot $t$ is $\hat{E}_m^{Uc}(t)=\rho_k(t)Y^SP_m^U(t)\hat{\tau}_m(t)$.

## IV. PROBLEM AND MDP FORMULATION

In this section, we first analyze the problem-solving difficulty and then introduce the HHGNN framework.

Subsequently, we model the global problem $\mathbb{GP}$, subproblems $\mathbb{L}1$, and $\mathbb{L}2$, analyze the feasibility of their decoupled solutions, and finally design the MDPs for $\mathbb{L}1$ and $\mathbb{L}2$.

### *A. HHGNN Framework*

The target of this work is to jointly optimize UAVs' trajectories $\boldsymbol{Q} = \{\boldsymbol{q}_m^U(t)\}$, WET decisions $\boldsymbol{Z} = \{Z_m(t)\}$, UAV-LEO scheduling $\boldsymbol{S} = \{s_m^{l^k}(t)\}$, UAV transmit power $\boldsymbol{P} = \{P_m^U(t)\}$, and SAT subchannels allocation $\boldsymbol{\rho} = \{\rho_k(t)\}$, with the aim to improve AoI of all GTs/UAVs, energy and bandwidth utilization efficiency and so on.

Classical combinatorial optimization or DRL methods may struggle to jointly optimize the five variables due to three critical limitations: 1) The dynamic path planning and limited coverage range of multiple UAVs, along with the multi-choice scheduling from multiple UAVs to multiple LEO SATs, result in a complex network topology. Coupled with the interdependencies among optimization variables, this renders the computational complexity of solution algorithms intractable. 2) The UAV's limited sensing range prevents a centralized controller from determining all UAVs' $\boldsymbol{Q}$, $\boldsymbol{Z}$, $\boldsymbol{S}$, and $\boldsymbol{P}$, while SAT subchannels $\boldsymbol{\rho}$ are contention-based resources among UAVs and cannot be jointly optimized with UAV characteristic variables in a decentralized manner. 3) As the PDPCIN framework divides the SAGIN architecture into G2A and A2S scenarios, UAV data transmission in the latter scenario unidirectionally depends on data collection in the former, such that action decisions in the former affect the environment states in the latter, making it impossible to use a single neural network for unified decision-making across all five variables.

Therefore, we propose the HHGNN framework using hierarchical decoupling: First, formulate the global optimization problem ($\mathbb{GP}$), then decompose it into two subproblems ($\mathbb{L}1$ and $\mathbb{L}2$) through the PDPCIN framework. Each subproblem is modeled as an MDP, with variables $\boldsymbol{Q}$, $\boldsymbol{Z}$, $\boldsymbol{S}$, $\boldsymbol{P}$, and $\boldsymbol{\rho}$ allocated to the respective MDPs. The PDPCIN topology is represented as a heterogeneous graph, where $\mathbb{L}1$ and $\mathbb{L}2$ are solved sequentially using multi-agent reinforcement learning (MARL) and single-agent reinforcement learning (SARL) architectures to achieve an indirect global solution. Moreover, by integrating the HHGNN framework, we further design S-LSDO algorithm to optimize SAT demand density. Detailed methodology follows.

### *B. Problem Formulation*

To improve the overall system performance, we need to clarify the data transmission and energy consumption of each layer. In G2A scenario, the total data volume transmitted from GTs to UAVs is $D^G(t) = \sum_{m=1}^{M} D_m^U(t)$, and the total energy consumption is $E(t) = \sum_{n=1}^{N} E_n^{Gc}(t) + \sum_{m=1}^{M} E_m^{Uc}(t)$ in slot $t$. Accordingly, in G2A scenario, the total transmission capacity and data volume transmitted from UAVs to LEO SATs are $D^U(t) = \sum_{m=1}^{M} \dot{D}_m^U(t)$ and $\hat{D}^U(t) = \sum_{m=1}^{M} \hat{D}_m^U(t)$, and the total energy consumption is $\hat{E}(t) = \sum_{m=1}^{M} \hat{E}_m^{Uc}(t)$ in slot $t$. Thus, the global optimization problem can be formulated as follows:

$$\mathbb{GP}: \max_{\boldsymbol{Q},\boldsymbol{Z},\boldsymbol{S},\boldsymbol{P},\boldsymbol{\rho}} \frac{\hat{D}^U(t)}{\beta\left(E(t)+\hat{E}(t)\right)\left(A^G(t)+A^U(t)\right)}, \tag{17}$$

$$s.t. \text{ (3), (7), (9), C1-C12,}$$

where $\beta$ is a positive scaling factor used to balance the magnitudes of denominator and numerator, constraint C1 to C12 will be introduced later in $\mathbb{L}1$ and $\mathbb{L}2$.

Problem $\mathbb{GP}$ involves multiple optimization variables and segmented data transmission, classifying it as a MINLP problem. Treating it as a monolithic optimization problem becomes computationally intractable. According to HHGNN Framework, we further decouple problem $\mathbb{GP}$ into two layered subproblems $\mathbb{L}1$ and $\mathbb{L}2$. The global solution is indirectly obtained through iterative subproblem resolutions. The formulation of the first layer optimization subproblem $\mathbb{L}1$ is

$$\mathbb{L}1: \max_{\boldsymbol{Q},\boldsymbol{Z}} \frac{D^G(t)}{\beta E(t) A^G(t)}, \tag{18}$$

$$s.t. \text{ (3), (7), (9)}$$

$$\| \boldsymbol{v}_m(t) \|_2 \leq v_{max}^U, \forall m \in \mathcal{V}^U, \tag{C1}$$

$$q_m^U(t) \in [0, L_E]^2, \forall m \in \mathcal{V}^U, \tag{C2}$$

$$\textstyle\sum_{m=1}^{M} s_{n,m}(t) \leq 1, \forall n \in \mathcal{V}^G, \tag{C3}$$

$$\textstyle\sum_{n=1}^{N} s_{n,m}(t) \leq Y^U, \forall m \in \mathcal{V}^U, \tag{C4}$$

$$B_m^U(t_{end}) \geq B_{min}^U, \forall m \in \mathcal{V}^U, \tag{C5}$$

$$d'_{m,i}(t) \geq d_{min}, \forall m, i \in \mathcal{V}^U, m \neq i, \tag{C6}$$

$$z_{min}^U(t) \leq z_m^U(t) \leq z_{max}^U, \forall m \in \mathcal{V}^U, \tag{C7}$$

$$\boldsymbol{C}(t) * \boldsymbol{S}(t) = \boldsymbol{S}(t). \tag{C8}$$

The goal of $\mathbb{L}1$ is to maximize transmission performance by optimizing $\boldsymbol{Q}$ and $\boldsymbol{Z}$ in G2A scenario, where operational constraints are defined by (C1–C7): (C1) UAV velocity $\boldsymbol{v}_m(t) \in \mathbb{R}^3$ is capped at $v_{max}^U$; (C2) UAVs are confined within the area of interest with border $L_E$; (C3) each GT is served by at most one UAV on a subchannel; (C4) the number of GTs assigned to each UAV does not exceed available subchannels; (C5) UAV battery energy $B_m^U(t_{end})$ must remain above $B_{min}^U$ for safe return at the end slot $t_{end}$; (C6) maintain a safe distance between UAVs; (C7) The flight altitude of UAVs is restricted within the range of $z_{min}^U(t)$ to $z_{max}^U$; (C8) UAVs can only server GTs within their covering range. Next, the second layer optimization subproblem $\mathbb{L}1$ is formulated as:

$$\mathbb{L}2: \max_{\boldsymbol{S},\boldsymbol{P},\boldsymbol{\rho}} \frac{D^U(t)}{\beta \hat{E}(t) A^U(t)}, \tag{19}$$

$$s.t. \textstyle\sum_{k\in\mathcal{K}} \rho_k \leq 1, \tag{C9}$$

$$0 < P_m^U(t) < P_{max}^U, \tag{C10}$$

$$\textstyle\sum_{k=1}^{K} \sum_{l^k \in \mathcal{V}^k} s_m^{l^k}(t) \leq 1, \forall m \in \mathcal{V}^U, \forall k \in \mathcal{K}, \tag{C11}$$

$$\hat{B}_m^U(t+1) = \min\left\{\begin{matrix} \hat{B}_m^U(t) + E_m^{Uh}(t) \\ -\hat{E}_m^{Uc}(t), \hat{B}_{max}^U \end{matrix}\right\}, \forall m \in \mathcal{V}^U. \tag{C12}$$

$\mathbb{L}2$ focuses on transmission performance improvement by optimizing $\boldsymbol{S}$, $\boldsymbol{P}$, and $\boldsymbol{\rho}$ in A2S scenario, where constraints (C9–C12) delineate the feasible operational domain: (C9) Sum of subchannel allocation ratios does not exceed 1; (C10) UAV transmit power is constrained within feasible range [0, $P_{max}^U$]; (C11) each UAV is restricted to connecting to exactly one LEO SAT at any given time; (C12) UAV energy storage from the harvesting board must not exceed its maximum capacity $\hat{B}_{max}^U$.

Next, we demonstrate that the combined solution of subproblems $\mathbb{L}1$ and $\mathbb{L}2$ constitutes the global optimum for $\mathbb{GP}$. For simplicity and without loss of generality, partition the variables into two disjoint sets: $\boldsymbol{x} = (\boldsymbol{Q}, \boldsymbol{Z})$ and $\boldsymbol{y} = (\boldsymbol{S}, \boldsymbol{P}, \boldsymbol{\rho})$. Let $\boldsymbol{x}^*$ and $\boldsymbol{y}^*$ denote the optimal solutions to $\mathbb{L}1$ and $\mathbb{L}2$, respectively. $(\boldsymbol{x}^*, \boldsymbol{y}^*)$ is optimal for $\mathbb{GP}$ under the consistency assumption: at optimality, the individual demands align $D_m^U(\boldsymbol{x}) = \dot{D}_m^U(\boldsymbol{y}), \forall m \in \mathcal{V}^U$. This consistency assumption implies that the data collected by each UAV from GTs is identical to that transmitted by each UAV to LEO SATs, ensuring no backlog or resource wastage, achieving supply-demand balance, and meeting the requirements of HHGNN framework optimal solution. Then, we obtain $\min\left(D_m^U(\boldsymbol{x}^*), \dot{D}_m^U(\boldsymbol{y}^*)\right) = D_m^U(\boldsymbol{x}^*) = \dot{D}_m^U(\boldsymbol{y}^*)$ , $\frac{D^G(x)}{\beta E(x)A^G(x)} \leq \frac{D^G(x^*)}{\beta E(x^*)A^G(x^*)} = v_1$, and $\frac{D^U(y)}{\beta \hat{E}(y)A^U(y)} \leq \frac{D^U(y^*)}{\beta \hat{E}(y^*)A^U(y^*)} = v_2$, where $v_1$ and $v_2$ are the optimal values of $\mathbb{L}1$ and $\mathbb{L}2$. Let $f_{GP}(\boldsymbol{x},\boldsymbol{y})$ denote the objective function of $\mathbb{GP}$. According to consistency assumption, inequality derivation, and tight bound theorem, we finally derive for any $(\boldsymbol{x},\boldsymbol{y})$ [41]:

$$f_{GP}(\boldsymbol{x},\boldsymbol{y}) \leq \min\left\{\frac{v_1 E(x)A^G(x)}{(E(x)+\hat{E}(y))\left(A^G(x)+A^U(y)\right)}, \frac{v_2 \hat{E}(y)A^U(y)}{(E(x)+\hat{E}(y))\left(A^G(x)+A^U(y)\right)}\right\}, \quad (20)$$

where the upper bound equals $f_{GP}(\boldsymbol{x}^*,\boldsymbol{y}^*)$ at $(\boldsymbol{x}^*,\boldsymbol{y}^*)$, and this value is attainable by the consistency assumptions, i.e., $f_{GP}(\boldsymbol{x},\boldsymbol{y}) \leq f_{GP}(\boldsymbol{x}^*,\boldsymbol{y}^*)$ . Therefore, the combination of the optimal solutions to $\mathbb{L}1$ and $\mathbb{L}2$ is the solution to $\mathbb{GP}$. However, owing to the non-convex nature of MINLP and the gradient descent optimization procedure, the $\mathbb{GP}$ solution derived herein is locally optimal. Nevertheless, it satisfies most application requirements at a lower cost compared to the pursuit of global optimality [42].

### *C. MDP Formulation*

#### *1) MARL MDP of* $\mathbb{L}1$

Based on the collaborative communication mode among UAVs, we formulate $\mathbb{L}1$ as a MARL task where each UAV is treated as an agent to observe environmental states and perform actions. Given the limited computational resources and observed environment states available to each UAV, we adopt a centralized training and distributed execution (CTDE) scheme to optimize the UAV training process. The following is the MDP formulation for a single UAV.

*a) Observation:* The observation of UAV-$m$ in slot $t$ encompasses its own features as well as those of surrounding GTs and UAVs located within its sensing range, described by associated neighborhood sets $\mathcal{N}_m^G(t) = \{n|g_{n,m}(t) = 1, \forall n \in \mathcal{V}^G\}$ and $\mathcal{N}_m^U(t) = \{i|u_{i,m}(t) = 1, \forall i \in \mathcal{V}^U\}$. Therefore, the observation of UAV-$m$ in slot $t$ is

$$o_m^t = \left\langle \begin{matrix} \boldsymbol{q}_m^U(t), B_m^U(t), \{\boldsymbol{q}_i^U(t), B_i^U(t)\}_{i\in\mathcal{N}_m^U(t)} \\ \{\boldsymbol{q}_n^G(t), B_n^G(t), \epsilon_n(t), a_n^G(t), \hat{D}_n^G(t)\}_{n\in\mathcal{N}_m^G(t)} \end{matrix} \right\rangle. \quad (21)$$

The attributes of UAVs include position and battery level, while those of GTs incorporate these two elements along with GT type, AoI, and cumulative data transmission volume, enabling UAVs to optimize trajectories and WET decision, accordingly. The observation length at each UAV varies across time slots and differs among UAVs, motivating the use of GNN-based solution.

*b) Action:* In each slot, each UAV-$m$ agent outputs UAV-$m$ 's velocity vector and WET decision, i.e., $a_m^t = \{\boldsymbol{V}_m(t), Z_m(t)\}$, and $\boldsymbol{q}_m^U(t+1)$ can be easily calculated based on UAV-$m$ 's three-dimensional velocity vector $\boldsymbol{V}_m(t) = \{V_m^X(x), V_m^Y(t), V_m^Z(t)\}$. Given that $\boldsymbol{V}_m(t)$ is continuous while $Z_m(t)$ are discrete, we employ neural networks with continuous action outputs and quantize the interval $[-1, 1]$ corresponding to $Z_m(t)$ for more precise trajectory optimization. Specifically, positive values map to $Z_m(t) = 1$ and negative values to $Z_m(t) = 0$.

*c) Reward:* Since UAVs can perform two behaviors, i.e., WDC and WET, we set up separate reward formulas for them. The reward for UAV-$m$ regarding WDC in slot $t$ is

$$r_m^{WDC}(t) = \sum_{n=1}^{N} a_n^G(t) W^U \log\left(1+\gamma_{n,m}(t)\right)\tau. \quad (22)$$

Formula (22) ensures that UAVs prioritize serving GTs with high AoI $a_n^G(t)$ while considering data collection. Accordingly, the reward for UAV-$m$ regarding WET in slot t is

$$r_m^{WET}(t) = \sum_{n=1}^{N} a_n^G(t) \frac{E_n^{Gh}(t) P_E^U Z_m(t) h_{n,m}(t)}{\sum_{m=1}^{M} P_E^U Z_m(t) h_{n,m}(t)}. \quad (23)$$

Since the energy harvesting contribution of each UAV to each GT varies, (23) requires calculating the contribution ratio $\frac{P_E^U Z_m(t) h_{n,m}(t)}{\sum_{m=1}^{M} P_E^U Z_m(t) h_{n,m}(t)}$ of UAV-$m$. Meanwhile, priority in energy provision is given to GTs with high AoI to ensure that they have sufficient energy to transmit data promptly. Combining (22) and (23), the total reward formula for UAV-$m$ is obtained as

$$r_m^t = \frac{\varsigma_1 r_m^{WDC}(t) + \varsigma_2 r_m^{WET}(t)}{A^G(t) E_m^{Uc}(t)}, \quad (24)$$

where $\varsigma_1$ and $\varsigma_2$ are non-negative coefficients to ensure that the distinct rewards maintain identical or comparable magnitudes. The rewards are divided by the total GT AoI $A^G(t)$ and UAV-$m$ 's energy consumption $E_m^{Uc}(t)$ to enable rewards to objectively reflect UAVs' contribution to transmission under different transmission environments, while driving UAVs to improve energy efficiency. Furthermore, no penalty terms for constraint violations are added because invalid actions are masked, which will be detailed in Section V.

#### *2) SARL MDP of* $\mathbb{L}2$

To avoid contention between LEO SATs for subchannels and additional communication delay caused by redundant collaborative communication between agents in A2S scenarios, we formulate $\mathbb{L}2$ as a SARL task. Specifically, the steps are as follows: arbitrarily select a LEO and deploy an agent model as the central controller in each of its SATs; during the training phase, when the coverage of the currently serving SAT is about to exclude the target area, the SAT transfers its agent parameters to the successor SAT, with continuous iteration; during the execution phase, since agent models no longer update, the agent parameters in all SATs are identical, eliminating the need for parameter transmission. The following is the MDP formulation.

*a) State:* The SAT agent's state space comprises UAVs' AoI, transmission task amount, energy harvesting board's battery level, UAV-SAT channel gain, and LEO SAT location $\boldsymbol{q}_{l^k}^S(t) = (x_{l^k}^S(t), y_{l^k}^S(t), z_{l^k}^S(t))$, which can be represented as

$$\mathcal{S}^t = \left\langle \begin{matrix} \{a_m^U(t)\}; \{\hat{D}_m^U(t)\}; \{\hat{B}_m^U(t)\}; \\ \left\{h_m^{l^k}(t)\right\}; \{\boldsymbol{q}_{l^k}^S(t)\} \end{matrix} \right\rangle_{\forall m\in\mathcal{V}^U, l^k\in\mathcal{V}^k, \forall k\in\mathcal{K}}. \quad (25)$$

*b) Action:* In each slot, the SAT agent makes decisions on UAV-LEO scheduling, UAV transmit power control, and LEO's subchannel allocation, with its action space defined as follows:

$$\mathcal{A}^t = \left\langle\left\{s_m^{l^k}(t)\right\}; \{P_m^U(t)\}; \{\rho_k(t)\}\right\rangle_{\forall m\in\mathcal{V}^U, l^k\in\mathcal{V}^k, \forall k\in\mathcal{K}}. \quad (26)$$

*c) Reward:* To enhance the actual data transmission volume of UAVs while reducing their AoI and energy consumption, the reward function is designed as follows:

$$\mathcal{R}^t = \sum_{m=1}^{M} \frac{a_m^U(\mathrm{t})}{A^U(\mathrm{t})} \frac{\widehat{D_m^U}(t)}{\widehat{E_m^{Uc}}(t)}. \qquad (27)$$

## V.NEURAL NETWORKS AND ALGORITHMS DESIGN

In this section, we define diverse GNN layers, construct the network architecture of HHGNN, design corresponding algorithms, and finally analyze the algorithms' time complexity.

### A. Neural Networks Design

*1) Heterogeneous Graphs Construction:* As shown in Fig. 2, we adopt a heterogeneous graph $\mathcal{G} = \langle \mathcal{V}, \mathcal{E} \rangle$ to characterize the device vertices of GTs, UAVs, and LEO SATs, as well as their mutual relations, where $\mathcal{V} = \mathcal{V}^G \cup \mathcal{V}^U \cup \{\mathcal{V}^k\}_{\forall k \in \mathcal{K}}$. The relations in $\mathcal{E}$ include: sensing: (GT, sensed-by, UAV), (UAV, sensed-by, UAV), (UAV, sensed-by, SAT), and (SAT, sensed-by, SAT); exchange: (UAV, exchange-with, UAV) and (SAT, exchange-with, SAT); service: (GT, transmit-to, UAV), (UAV, power, GT), and (UAV, transmit-to, SAT). A sensing relation $(i, j)$ exists if device $i$ is within $j$'s sensing or coverage range. An exchange relation exists if UAVs are mutually within each other's sensing range. A service relation $(i, j)$ exists on the premise that device $i$ is within $j$'s coverage range. The sensing or exchange relations between SATs exist only if they are currently covering the target area in each orbit. Additionally, since GTs harvest all RF power from UAVs and perform non-linear power transformation, UAVs conducting WET have service relations with all E-GTs. It is noted that Fig. 2 reflects the actual inter-device relations. In GNNs, only UAVs or SATs serve as destination vertices for state information transmission (rather than GTs), as this aligns with real-world communication scenarios where high-load tasks are not processed on GTs with limited computational resources. As observed from Fig. 2, the overall heterogeneous graph comprises multiple local sensing and exchange subgraphs. Next, we define GSL and GEL to manage sensing and exchange features, and define GML to handle unavailable actions output by the GNN.

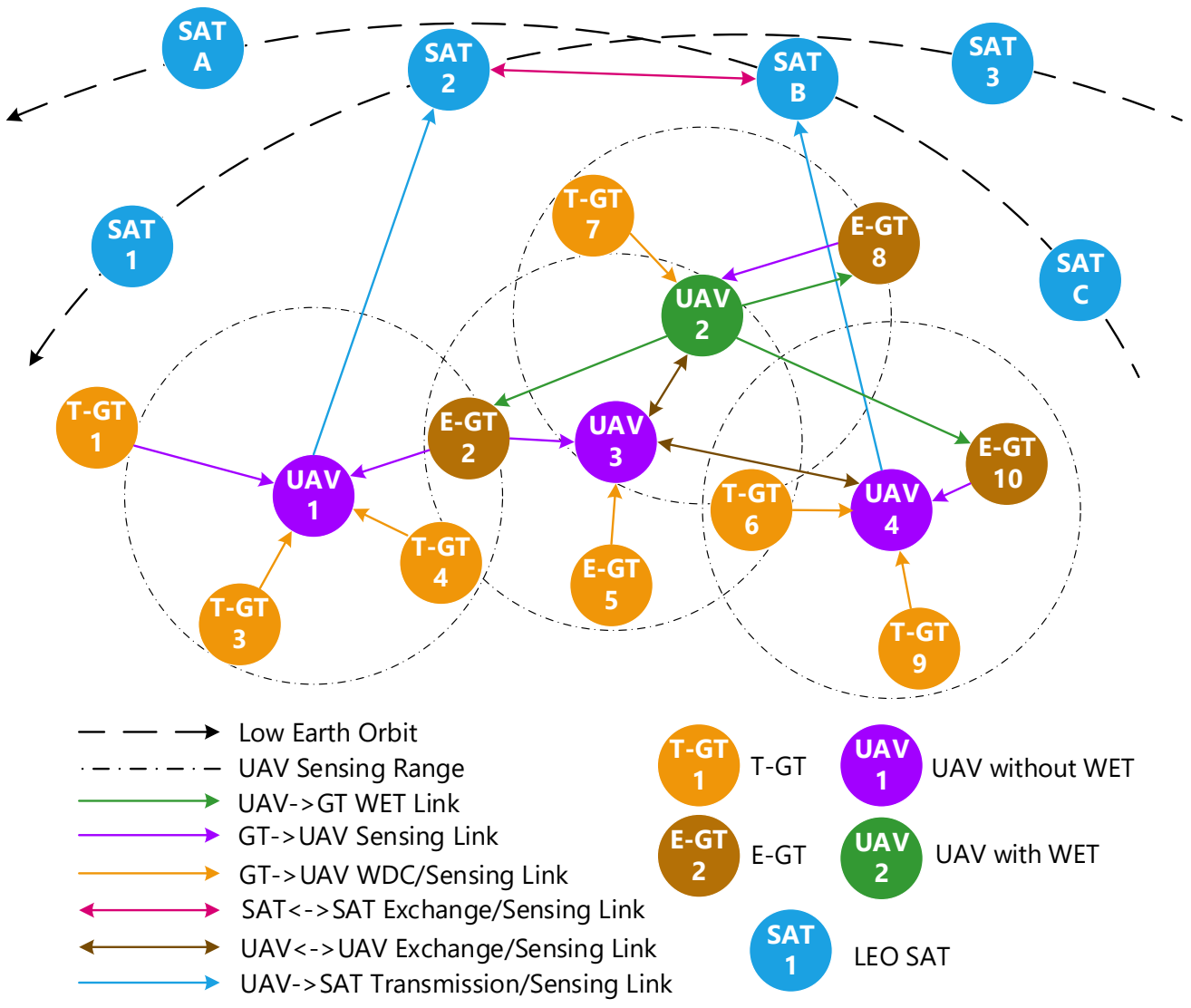

Fig. 2. Graph representation of PDPCIN topology.

*2) Graph Sensing Layer:* Most existing works encode features by a multi-layer perceptron (MLP), which can only process intrinsic vertex features and fail to effectively handle graph-structured features due to their inability to aggregate neighbor information. In contrast, graph neural networks (GNNs) capture topological relations and vertex dependencies via message passing, making them more suitable for the time-varying concatenated communication scenario in this study. Given the limited expressive power of traditional graph attention (GAT) with static attention mechanisms, we adopt strictly more expressive GATv2 [43]. Since the G2A scenario observations include GTs and UAVs, while the A2S scenario states involve UAVs and SATs, we employ two GATv2 in each GSL to weight information from different source vertices at each destination vertex. The weight coefficient for each GT, UAV, and SAT is obtained as

$$\begin{cases} \alpha_{in}^G = \dfrac{\exp\left(\mathrm{LeakyReLU}\left(\mathbf{a}_G^{\mathrm{T}}(\mathbf{W}_{agent}\mathbf{h}_i || \mathbf{W}^G \boldsymbol{v}_n^G(t))\right)\right)}{\sum_{l \in \mathcal{N}_m^G(t)} \exp\left(\mathrm{LeakyReLU}\left(\mathbf{a}_G^{\mathrm{T}}(\mathbf{W}_{agent}\mathbf{h}_i || \mathbf{W}^G \boldsymbol{v}_l^G(t))\right)\right)}, \\ \alpha_{im}^U = \dfrac{\exp\left(\mathrm{LeakyReLU}\left(\mathbf{a}_U^{\mathrm{T}}(\mathbf{W}_{agent}\mathbf{h}_i || \mathbf{W}^U \boldsymbol{v}_m^U(t))\right)\right)}{\sum_{l \in \mathcal{N}_m^U(t)} \exp\left(\mathrm{LeakyReLU}\left(\mathbf{a}_U^{\mathrm{T}}(\mathbf{W}_{agent}\mathbf{h}_i || \mathbf{W}^U \boldsymbol{v}_l^U(t))\right)\right)}, \\ \alpha_{il^k}^S = \dfrac{\exp\left(\mathrm{LeakyReLU}\left(\mathbf{a}_S^{\mathrm{T}}\left(\mathbf{W}_{agent}\mathbf{h}_i || \mathbf{W}^S \boldsymbol{v}_{l^k}^S(t)\right)\right)\right)}{\sum_{l \in \mathcal{N}^S(t)} \exp\left(\mathrm{LeakyReLU}\left(\mathbf{a}_S^{\mathrm{T}}(\mathbf{W}_{agent}\mathbf{h}_i || \mathbf{W}^S \boldsymbol{v}_l^S(t))\right)\right)}, \end{cases} \qquad (28)$$

where $\mathbf{h}_i$ is agent-$i$'s current features, $\mathcal{N}^S(t)$ is the set of SATs currently covering the target area in each orbit, $\boldsymbol{v}_n^G(t)$, $\boldsymbol{v}_m^U(t)$, $\boldsymbol{v}_{l^k}^S(t)$ is the initial features of GT-$n$, UAV-$m$, and SAT-$l^k$, $\mathbf{a}^{\mathrm{T}}$ is $\mathbf{a}$'s transpose matrix, $||$ represents concatenation, and $\mathbf{W}_{agent}$, $\mathbf{W}^G$, $\mathbf{W}^U$, $\mathbf{W}^S$, $\mathbf{a}_G$, $\mathbf{a}_U$, and $\mathbf{a}_S$ are trainable vectors. The updated features, derived as the weighted sum of features from all incident neighbors, is calculated through

$$\begin{cases} \mathbf{h}_i^G := \sigma\left(\sum_{l \in \mathcal{N}_m^G(t)} \alpha_{in}^G \, \mathbf{W}^G \boldsymbol{v}_l^G(t)\right), \\ \mathbf{h}_i^U := \sigma\left(\sum_{l \in \mathcal{N}_m^U(t)} \alpha_{im}^U \, \mathbf{W}^U \boldsymbol{v}_l^U(t)\right), \\ \mathbf{h}_i^S := \sigma\left(\sum_{l \in \mathcal{N}^S(t)} \alpha_{il^k}^S \, \mathbf{W}^S \boldsymbol{v}_l^S(t)\right), \end{cases} \qquad (29)$$

where $\sigma$ is the sigmoid activation function. To enhance training stability, multi-head attention is applied by independently deploying multiple attention mechanisms, whose outputs are concatenated. The observation embedding for GTs is computed as $\mathbf{h}_i^G := ||_{a=1}^{A} \sigma\left(\sum_{l \in \mathcal{N}_i^G(t)} \alpha_{in}^a \, \mathbf{W}_a^G \boldsymbol{v}_l^G(t)\right)$ where $A$ is the number of heads. The embeddings of $\mathbf{h}_i^U$ and $\mathbf{h}_i^S$ can be obtained similarly. These are then fused via an MLP to produce the final vertex representation $\mathbf{h}_i$. The self-attention mechanism enables GATv2 to selectively process inputs from dynamically varying sets of GTs/UAVs/SATs based on relevance. The heterogeneous GATv2 layer dedicated to features sensing is termed graph sensing layer (GSL).

*3) Graph Exchange Layer:* While GAT performs well in sensing features processing, (29) reliance on exchanging high-dimensional hidden features makes it inefficient for inter-agent exchange. As the feature dimension increases, so does the exchange cost. To reduce backhaul burden, feature compression is required. Existing methods either treat discrete messages as agent actions, requiring auxiliary reinforcement learning due to the absence of gradients, or transmit continuous messages with quantization noise [44].

In this work, we adopt a differentiable discrete messaging scheme, and define the GEL with an encoder-decoder architecture. Specifically, edge update function $\rho^e_{exc}$, parameterized by MLP $N_{enc}$, works as an encoder and transforms UAV/SAT features into logits over a discrete symbol set, which are subsequently utilized to select symbols for transmission. To enable backpropagation, we use the Gumbel-Softmax trick, which reparameterizes categorical sampling into a differentiable form. Gumbel-Softmax approximates a one-hot vector as temperature $\tau^e \to 0$, allowing gradient flow during training, and hard symbols are sampled directly during inference [29]. In detail, function $\rho^e_{exc}$ first aggregates discrete messages via a max-pooling operation at the receiver side, then $N_{dec}$ decodes the outcome. Recurrent neural network (RNN), working as vertex update function, is used to take current vertex features and aggregated information from neighboring agents. Thus, the update of UAV features is expressed as

$$\mathbf{h}_i := \mathrm{RNN}\left[\mathbf{h}_i || N_{dec}\left(\max_{l\in\mathcal{N}^U_m(t)}\left(\mathrm{GS}\left(N_{enc}(\mathbf{h}_l)\right)\right)\right)\right], \quad (30)$$

where GS is abbreviation for Gumbel-Softmax. SAT features can be processed in the same way.

*4) Graph Mask Layer:* As indicated by constraints (C1-C12), the optimization problems are subject to multiple restrictions. To ensure neural network outputs comply with these constraints, two common strategies, soft-constraint and hard-constraint, are available. Soft-constraint discourages invalid actions by assigning low or negative rewards but cannot fully prevent them, while hard-constraint preemptively masks invalid actions, ensuring feasibility but requiring prior knowledge of all infeasible actions. Since some constraints in our scenario are only verifiable post-decision, hard constraints are inapplicable. Therefore, we propose a new constraint method, GML, based on the characteristics of GNN structure. It determines the action vector $\mathbf{a}$ based on the graph features $\mathbf{h}_i$. If actions are feasible, they are used directly. Otherwise, use smooth-mask mechanism to adjust the distribution of actions, ensuring actions align with the graph structure and remain valid. The modification of $\mathbf{a}$ is

$$\mathbf{a}' = \begin{cases} \mathbf{a}, & \text{if } f_{\mathcal{G}}(\mathbf{h}_i, \mathbf{a}) \in \Omega_{\mathcal{G}} \\ f_{\Omega_{\mathcal{G}}}\left(\mathcal{G}_\theta(\mathbf{a}) \circ (\boldsymbol{\mu}_\theta + \boldsymbol{\sigma}_\theta + \varepsilon)\right), & \text{otherwise} \end{cases}, \quad (31)$$

where $\varepsilon$ is noise, $f_{\mathcal{G}}(\cdot)$ is global risk assessment function evaluating whether the state-action pair is within the feasible domain $\Omega_{\mathcal{G}}$, $\boldsymbol{\mu}_\theta$ and $\boldsymbol{\sigma}_\theta$ are the mean and standard deviation vector of $\mathbf{a}$ output by MLP, $\mathcal{G}_\theta(\cdot)$ is a smooth function used to dynamically adjust the feasibility weights of each dimension of the action, $\circ$ is element-wise multiplication combining the GNN's smooth vector with the original action distribution to smooth the discrepancy between infeasible and normal actions, $f_{\Omega_{\mathcal{G}}}(\cdot)$ is the projection operator to fine-tune infeasible actions via noise addition to keep them within feasible domain $\Omega_{\mathcal{G}}$. The proposed method fully blocks infeasible actions while balancing the rationality of corrected actions to maximally preserve their original features.

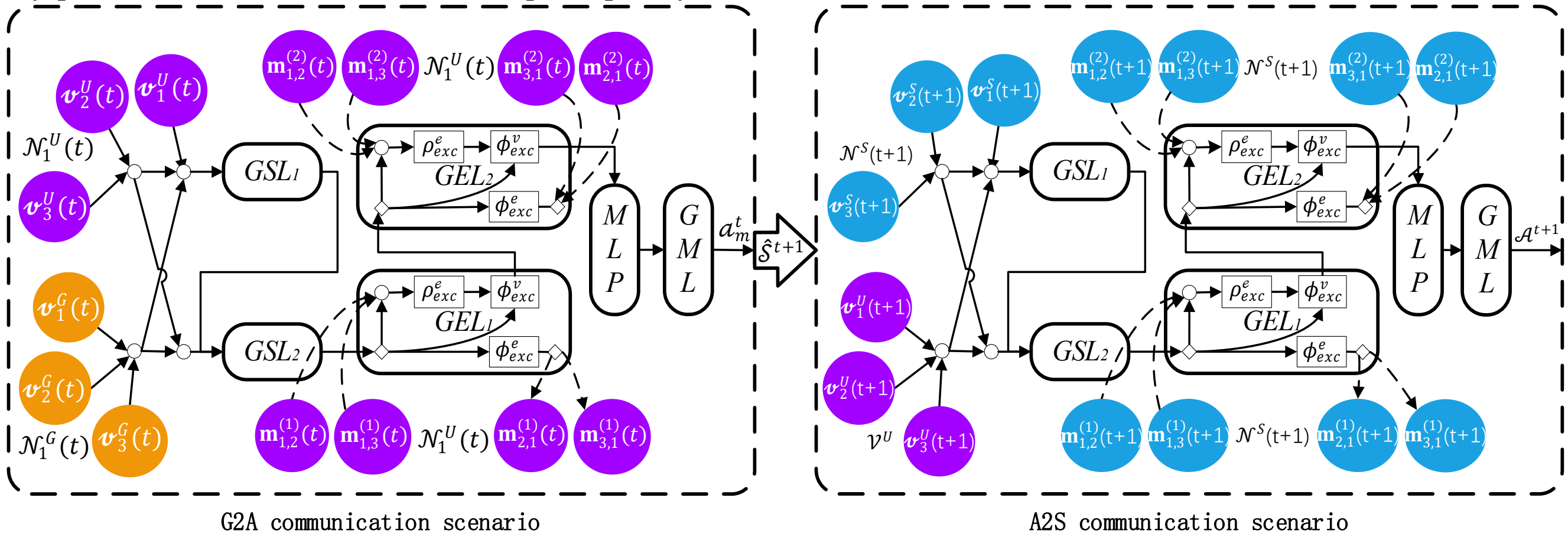

Fig. 3. HHGNN architecture with G3Ms. Sensing, exchange, and masking correspond to agent-1 in G2A and A2S scenario.

*5) Concatenated Structure:* As show in Fig. 3, multiple layers are stacked to enhance model capacity, forming a concatenated structure, where $\mathbf{m}^{(l)}_{i,j}(t)$ denotes the message of neighbor vertex-$j$ updated by the $l$-th GEL after processing via GSLs, $\phi^v_{exc}$ and $\phi^e_{exc}$ are vertex update and edge update functions, and the updated state $\hat{\mathcal{S}}^{t+1} = \langle\{a^U_m(t+1)\}; \{\hat{D}^U_m(t+1)\}; \{\hat{B}^U_m(t+1)\}\rangle_{\forall m\in\mathcal{V}^U}$ after all UAVs execute decisions in slot $t$ within the G2A scenario constitutes a part of the MDP's state $\mathcal{S}^{t+1}$ in A2S scenario. Let $L_S$ and $L_E$ denote the number of GSL and GEL, respectively. In slot $t$, agent-$i$'s features after $l$-th update are denoted $\mathbf{h}^{(l)}_i(t)$, where $0 \le l \le L_S + L_E$. The neural network model incorporating GSL, GEL, and GML is named G3M for briefness. Fig. 3 only shows one G3M in G2A scenario, which involves multiple G3Ms in fact, with each UAV having a G3M as an agent. Taking G2A scenario as an example, during forward propagation, GSL update only destination vertices, leaving source vertex features unchanged. Thus, when agent-$i$ is the destination, its feature evolves as $\mathbf{h}^{(l)}_i(t)$; when it is a source (i.e., observed by others), its features remain $\mathbf{h}^{(0)}_i(t)$. After applying $L_S$ GSL locally, agent-$i$ obtains hidden features $\mathbf{h}^{(L_S)}_i(t)$, which is further refined through $L_E$ GEL via message exchange with neighbors, i.e., $\mathcal{N}^U_m$ or $\mathcal{N}^S(t)$. If an agent has no neighbors, a zero vector is used as the aggregated message. Notably, features from neighbors contain their own features in GSL, while integrated messages from neighbors include not only their own feature information but also information received from other vertices and some historical information in GEL.

### B. Algorithms Design

*1) Training Process of G3M:* $\mathbb{L}1$ and $\mathbb{L}2$ are addressed using CTDE-MARL and SARL, respectively, which rely on environmental interactions for training through dynamic state-action-reward data collection. Both aim to maximize cumulative rewards via classic RL frameworks (e.g., policy gradient and value function estimation), following an iterative data sampling-evaluation-optimization loop to adapt strategies through environmental feedback. Therefore, taking the G2A scenario as an example, we introduce the training process of G3M. For the training procedures in the A2S scenario, the number of agents can be set to 1, and the corresponding environmental parameters should be replaced accordingly. The following are the specific steps of Algorithm 1.

**Algorithm 1:** Training Process of G3M

```
for episode from 1 to N^{ep} do
    Reset the environment and devices states;
    while t ≤ T do
        for all agent i ∈ V^U do
            Get access to N_m^G(t) and N_m^U(t) to receive o_i^t;
            for l from 1 to L_S do
                Apply GSL on local graph and obtain features h_i^{(l)}(t);
            end for
            for l from 1 to L_E do
                Encode and send messages m_{ji}^{(l)}(t) to j ∈ N_m^U(t) via GEL;
                Decode and aggregate messages m_{ij}^{(l)}(t) from j ∈ N_m^U(t)
                    with hidden state z_i^t via GEL, and obtain z_i^{t+1};
            end for
            Obtain a_i^t via GML, receive reward r_i^t, and observe o_i^{t+1};
        end for
        Store ⟨{o_i^t, z_i^t, a_i^t, r_i^t, o_i^{t+1}, z_i^{t+1}}_{i∈V^U}⟩ into replay memory D;
        If |D| > |B| then
            Draw a batch of samples B and update model weights θ by
                classic loss function;
        End if
        Update weights of target network θ̂ by θ̂ = ξθ + (1 − ξ)θ̂;
        t ← t + 1;
    End while
End for
```

Each agent's computational complexity is determined by the forward computation of the trained network, which comprises four components: GSL for sensing observation encoding, GEL for multi-vertex exchange, MLP for feature forward propagation, and GML for unavailable action masking. Their corresponding time complexities are $O_1 = \mathcal{O}(L_S(AN_{nei}d_{\text{in}}d_{\text{out}} + ANd_{\text{out}}))$ , $O_2 = \mathcal{O}(L_E N_{nei} d_{\text{in}} d_{\text{out}})$ , $O_3 = \mathcal{O}\left(\sum_{j=2}^{L_A} N'_{j-1} \cdot N'_j\right)$ and $O_4 = \mathcal{O}(dim(\boldsymbol{a}))$ [6], [45], where $N_{nei}$ is the number of neighbors, $d_{\text{in}}$ and $d_{\text{out}}$ are input and output feature dimensions, $L_A$ and $N'_j$ are the number of fully connected (FC) layers of MLP and the number of neurons in the $j$-th layer, and $dim(\boldsymbol{a})$ is the dimension of action $\boldsymbol{a}$. Thus, the total time complexity of G3M is approximated by $O^{G3M} = \sum_{i=1}^{4} O_i N^{ep} T$ . Since complexity scales with the number of neighbors, neurons, FC layers, and action dimension, the model flexibly accommodates varying environmental carrying capacity and communication service demand.

*2) S-LSDO Algorithm under HHGNN Architecture:* Higher LEO SAT density reduces UAV service intervals and S-AoI but increases deployment costs, inter-SAT interference, and channel contention. Conversely, lower density reduces interference and improves spectrum utilization at the cost of higher S-AoI due to coverage gaps. Optimizing this balance enables joint spectral efficiency-AoI enhancement, offering theoretical insights for LEO network design. Therefore, we propose the S-LSDO algorithm under HHGNN architecture to explore the trade-off between LEO SAT density, AoI, and spectrum utilization, as illustrated in Algorithm 2.

**Algorithm 2:** S-LSDO Algorithm under HHGNN Architecture

```
Input: S-AoI proportion range ∂^S, number of subchannels Y^S;
Output: the number of SATs in a single-LEO L^S;
Dispatch UAVs to the post-disaster area and allocate a G3M to each of
    them forming model set ℂ;
Find the SAT in each LEO currently providing services and label their
    successors as the first LEO SAT in each LEO;
Build the SAT set V^k, ∀k ∈ K of each LEO based on SAT motion
    patterns and allocate a G3M m to them;
Run Algorithm 1 to train model set ℂ for solving subproblem 𝕃1;
Initialize L^{max} with an empirical large value and L^{min} an empirical
    small value;
repeat
    N^S ← L^{min} + (L^{max} + L^{min} + 1)//2;
    Run Algorithm 1 using Ŝ^{t+1} (generated by model set ℂ) to train
        model m for solving subproblem 𝕃2;
    Invoke model m to get S-AoI proportion ∂;
    if ∂ > ∂^S then
        low ← L^S + 1;
    else if ∂ < ∂^S then
        high ← L^S − 1;
    end if
until ∂ in ∂^S
return L^S
```

The time complexity of Algorithm 2, composed of the number of searches and $\mathbb{L}2$ 's G3M iterative training, is expressed as $\mathrm{O}' = \mathcal{O}(\log(L^{max} - L^{min}) * O^{G3M})$ . The efficiency of the search stems from its logarithmic time complexity, making it suitable for fast retrieval in large-scale SATs. Suitable value of $L^{max}$ and $L^{min}$ will enhance the algorithm's efficiency. For this reason, we present the derivation of S-AoI proportion expected value in Section VI, which is intended to facilitate the selection of $L^{max}$ and $L^{min}$.

## VI. SIMULATION EXPERIMENTS

In this section, we first configure the experimental environment and key parameters, then present the relevant comparison schemes, subsequently conduct an analysis of the experimental results, and ultimately derive the expected value expressions for AoI and S-AoI proportion.

### A. Simulation Setups

Four UAVs are deployed to a 1.5 km × 1.5 km post-disaster area with 9 GTs randomly positioned. UAVs harvest up to 80% of their maximum energy harvesting panel capacity per time slot. Four LEO with a radius of 6921 km cover the area, referring to the coverage model in [25] and [46]. Each LEO's SAT configuration refers to Starlink Block v1.5, where SATs' number, altitude, and velocity are 22, 550 km, and 7.59 km/s, respectively.

In each G3M, we set $L_S = 2$ and $L_E = 2$ for GSL and GEL, respectively, balancing computational efficiency and performance. A single GML processes unavailable actions, while each GSL employs 4 attention heads. To further explore scheme performance in air-ground environments, an additional identical post-disaster area is added, deploying another 4 UAVs, creating an 8-UAV high-load A2S scenario to highlight performance differences and robustness. Other key parameters are listed in Table II.

TABLE II: SIMULATION PARAMETERS

| Parameter | Value | Parameter | Value |
|---|---|---|---|
| $\mathbb{N}_0$ | -174 dBm/Hz | $W^U, W^S$ | 1 MHz, 1 MHz |
| $z_{min}^U(t), z_{max}^U(t)$ | 60 m,120 m | $Y^U, Y^S$ | 2, 10-40 |
| $\alpha_L, \alpha_N, a, b$ | 3, 5, 12.08, 0.11 | $P^G, P_E^U, P_C^U$ | 10 mW,1 W, 10 mW |
| $P_{\text{sen}}, P_{\text{sat}}$ | -10 dBm, 7 dBm | $B_E, B_I, P_{max}^U$ | 0.01 J, 0.5 J, 1 W |
| $B_{max}^G, B_{max}^U, \hat{B}_{max}^U$ | 1 J, 500-2000 J, 10 J | $d_{min}, \tau$ | 10 m, 1 s |
| $O_S^G, O_S^U, O_C^U$ | 400 m, 400 m, 200 m | $v_{max}^U, V^S$ | 30 m/s, 7.59 km/s |
| $G_m^{lk}/T_N, \lambda_m^{lk}$ | 34 dB/K, 8 dB | $\mu_r, \sigma_r^2,$ | -2.6 dB, 1.63 dB |
| $\kappa$ | 1.38×10e-23 J/m | $d_E, h_k$ | 6371 km, 550 km |

*B. Comparison Schemes*

To further validate the effectiveness of our schemes IS-UAV and DMLA, and investigate the impact of optimization factors on the optimization problems, we develop four comparison schemes for $\mathbb{L}1$ and $\mathbb{L}2$, respectively.

*1) $\mathbb{L}1$ Comparison Schemes: a) DC-UAV:* This scheme is proposed in [6], where UAVs dynamically determine whether to perform WET or WDC. *b) TD-UAV:* This scheme, proposed in [18], divides UAVs into two teams: one solely responsible for WET and the other exclusively for WDC. *c) PD-UAV:* This scheme adopts the classic time phase division, where all UAVs perform WET before time slot $t'$ and WDC thereafter, with $t'$ determined via a greedy algorithm. *d) O-UAV:* This scheme deploys trajectory-fixed opportunistic UAVs following [47], with WDC/WET decisions operating under the PD-UAV strategy.

*2) $\mathbb{L}2$ Comparison Schemes: a) Frequency Division Multiple Access with Power Control (FDPC):* Neural networks optimize UAV-LEO scheduling, subchannel allocation, and UAV power control, with UAVs employing FDMA on their own dedicated subchannels. *b) Time Division Multiple Access with Fixed Control (TDFP):* This scheme partitions each time slot into equal segments based on the number of UAVs, where each UAV occupies a segment and transmits data to the nearest LEO SAT among all LEOs' SAT with fixed power across all subchannels. *c) Frequency and Time Division Multiple Access with Power Control (FTPC):* Unlike FDPC, this scheme uses FDMA for UAVs connected to different LEO SATs and TDMA for those connected to the same LEO SAT. *d) Uniform Allocation with Fixed Power (UAFP):* Subchannels are uniformly allocated to each LEO, with an equal number of fixed-power UAVs accessing each LEO's SATs.

*3) $\mathbb{GP}$ Comparison Schemes:* Our approach to solving $\mathbb{GP}$ iteratively employs solutions to $\mathbb{L}1$ and $\mathbb{L}2$, hence termed IS-UAV-DMLA. Moreover, given the superior performance of IS-UAV over $\mathbb{L}1$ comparison schemes (as demonstrated later), the comparative analysis for $\mathbb{GP}$ and single-LEO SAT density optimization integrates IS-UAV with $\mathbb{L}2$ comparison schemes, denoted as IS-UAV-FDPC, IS-UAV-TDFP, IS-UAV-FTPC, and IS-UAV-UAFP. In addition, we define two metrics to further evaluate transmission performance, i.e., energy transfer efficiency (ETE), which is the ratio of data transmission volume to energy consumption per unit time, and spectrum transmission efficiency (STE), which is the ratio of data transmission volume to bandwidth per unit time.

*C. Simulation Results*

*1) Performance Comparison of Different Schemes to* $\mathbb{L}1$

Fig.4(*a*) and Fig.4(*b*) depict the convergence behavior of data transmission and AoI across various schemes during the training phase. The proposed IS-UAV scheme achieves the most favorable performance, characterized by minimal fluctuations, followed by DC-UAV. Both IS-UAV and DC-UAV demonstrate greater adaptability in WDC/WET decision-making, enabling more effective state-action pair matching and superior convergence properties. In contrast, PD-UAV and TD-UAV exhibit comparable performance with relatively larger variations. These two schemes employ static partitioning strategies for WDC/WET decisions—either along the device or time dimensions. However, the extended operational cycle of TD-UAV introduces higher system instability, resulting in more pronounced training variability. O-UAV, which utilizes idle resources from other UAVs without relying on DRL-based decision-making, remains unaffected in terms of transmission performance during the training process.

Fig.4(*c*), Fig.4(*d*), and Fig.4(*e*) present a comparative analysis of average data transmission volume per UAV task cycle, AoI, and fairness among different schemes under varying UAV battery capacities ($B_{max}^U$). Overall, as $B_{max}^U$ increases, all schemes demonstrate improved transmission capabilities due to the availability of additional energy resources. Nevertheless, as shown in Fig.4(*d*), the rate of improvement diminishes with increasing battery capacity, and the AoI differences between schemes become less pronounced. This phenomenon occurs because energy saturation reduces the effectiveness of each scheme's distinct WDC/WET decision mechanism, making spectral bandwidth and path planning the dominant limiting factors. As illustrated in Fig.4(*e*), there is a slight decline in fairness across all schemes, primarily attributed to the heterogeneity in GT states and their respective transmission requirements, which prevent perfect service equity. Nonetheless, the observed level of fairness remains within an acceptable range.

Further comparative analysis reveals that the baseline O-UAV performs the worst but approaches the performance levels of PD-UAV and TD-UAV when $B_{max}^U$ is sufficiently large. TD-UAV and PD-UAV are outperformed by IS-UAV and DC-UAV due to their reliance on fixed resource partitioning strategies. Specifically, TD-UAV employs team-based execution of WDC or WET, leading to imbalanced workload distribution between UAV teams and lacking the ability to dynamically adjust team sizes based on real-time energy replenishment or task transmission needs. Meanwhile, PD-UAV executes WET and WDC in separate phases, causing significant data backlog during the early stages, which results in a sharp increase in AoI and negatively impacts information timeliness. This explains its inferior AoI performance compared to TD-UAV. In contrast, IS-UAV and DC-UAV dynamically allocate resources for WDC/WET operations, avoiding both idle and overloaded team issues and preventing AoI surges caused by pre-charging GTs. Among these, IS-UAV outperforms DC-UAV due to the latter's slot-level state switching mechanism, which inherently operates in a sequential processing mode, leading to inefficient utilization of time resources.

In practical applications, task switching incurs non-negligible delays—for example, reconfiguring communication parameters—which prolongs overall task completion time and fails to simultaneously address the dual urgent demands of data transmission and energy replenishment from GTs in post-disaster areas. For instance, when a GT with critically low battery power needs to transmit high-priority data, DC-UAV may not respond effectively. By contrast, IS-UAV enables parallel processing, allowing simultaneous WDC and WET within the same time window, thereby ensuring continuous data communication and sustainable energy supply for GTs. This eliminates potential device shutdowns or data loss caused by time-division waiting. Ideally, IS-UAV can reduce task completion time to less than 50% of that required by traditional time-division schemes, highlighting its significant efficiency gains and strong adaptability in handling dynamic and time-sensitive post-disaster communication scenarios.

Finally, Fig.4(*f*) shows the ETE statistics across all schemes and UAV battery capacities. It is evident that IS-UAV delivers the best overall performance, while O-UAV exhibits the lowest yet most stable ETE. Despite its poor overall performance, O-UAV achieves further enhancement in ETE by utilizing idle resources to provide supplementary communication services. Due to its phased and fixed WDC/WET operation mode, PD-UAV fails to meet dynamic environmental demands, resulting in the most significant fluctuations in ETE performance.

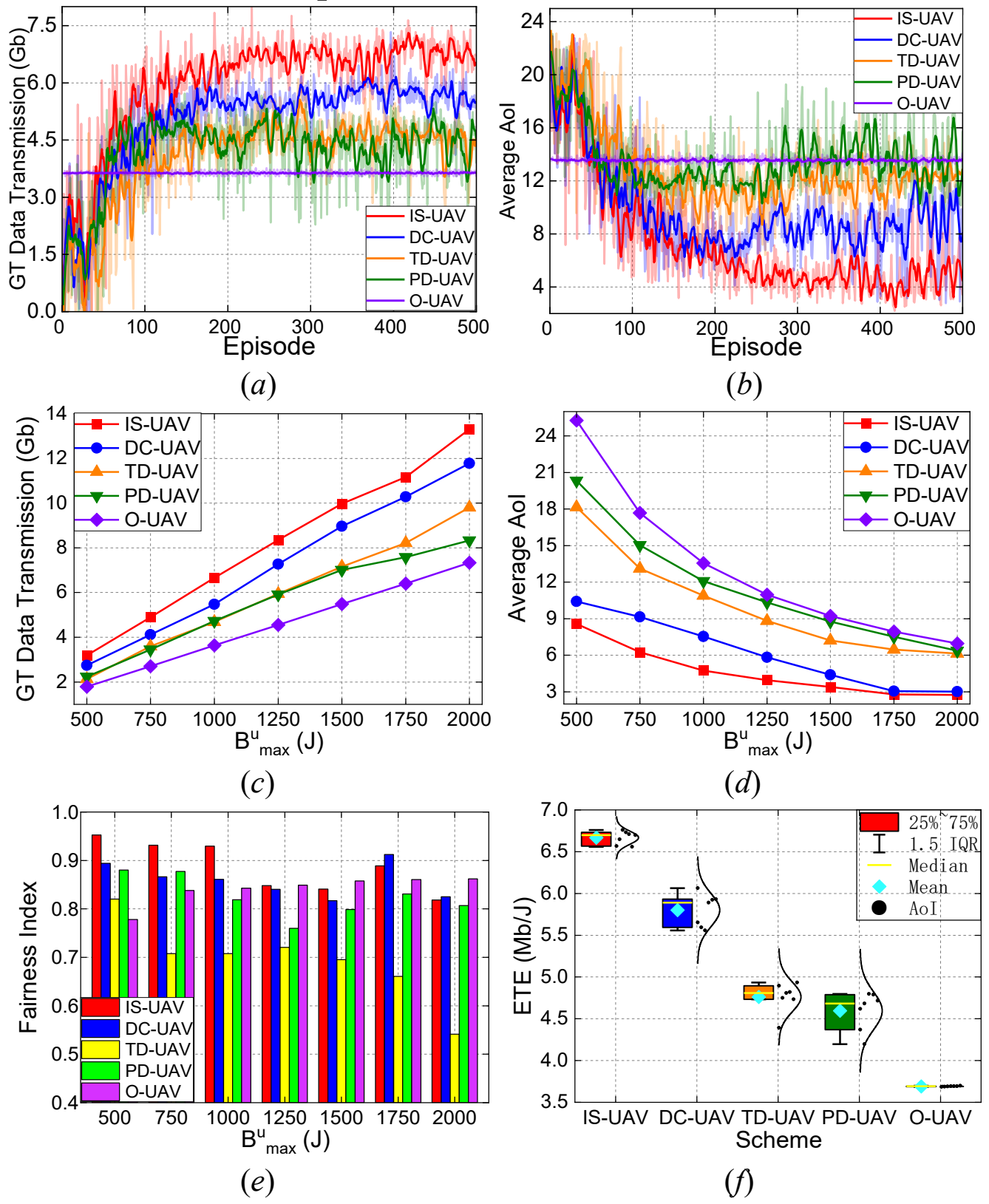

Fig. 4. Performance comparison of different schemes to 𝕃1 under different UAV battery capacities.

### 2) *Performance Comparison of Different Solutions to* 𝕃2

Fig.5(*a*) and Fig.5(*b*) show the average data transmission volume per time slot and STE of each scheme under varying numbers of subchannels. DMLA achieves the highest transmission performance, confirming that NOMA offers superior channel utilization efficiency compared to other access techniques. FTPC outperforms FDPC due to its implementation of fine-grained time division, which better adapts to dynamic channel conditions while mitigating interference caused by concurrent transmissions. TDFP and UAFP demonstrate nearly equivalent performance since both adopt uniform resource allocation strategies; however, TDFP slightly surpasses UAFP because each UAV in TDFP connects to the nearest SAT among all LEOs, thereby optimizing channel gain. In contrast, UAFP distributes resources evenly across the nearest SAT for each LEO, leading to suboptimal channel conditions. Nevertheless, TDFP's single-SAT access mechanism results in load imbalance among LEO SATs, whereas EDFP ensures the most balanced SAT load distribution across all schemes. It is also observed that the STE of all schemes declines with an increasing number of subchannels: when the total transmission demand remains constant, a larger number of subchannels combined with limited UAV energy storage leads to reduced power allocation per channel, lowering bandwidth utilization efficiency. DMLA and FTPC exhibit a more significant decrease in the STE as they prioritize balancing ETE and STE.

Fig.5(*c*) and Fig.5(*d*) display the average UAV energy consumption per time slot and the corresponding ETE under different subchannel configurations. Initially, when subchannels are limited, DMLA consumes significantly more energy than other methods and exhibits lower ETE. This is attributed to NOMA's ability to allow UAVs to share channels and allocate power across multiple subchannels, which, under resource constraints, necessitates increased transmit power to meet transmission demands. However, as the number of subchannels increases, the ETE of DMLA and FTPC improves. Benefiting from their strong transmission capabilities, these two schemes can reduce transmit power appropriately when bandwidth is sufficient, achieving a balance between ETE and STE, which explains their relatively stable energy consumption trends. FDPC demonstrates the best and most consistent ETE performance, as each UAV is assigned exclusive subchannels, enabling stable data transmission in every time slot. Furthermore, through integrated power control mechanisms, FDPC can adjust transmit power based on factors such as the number of allocated channels, individual transmission tasks, and channel quality.

Fig.5(*e*) and Fig.5(*f*) depict AoI and outage probability, where the latter is estimated using the ratio of outage signals to total signals according to the law of large numbers. The results indicate an inverse relationship between AoI and the previously analyzed transmission capacity of each scheme. As the number of subchannels increases, the rate of AoI reduction gradually diminishes or even plateaus, suggesting that bandwidth saturation occurs and energy availability becomes the primary constraint. These findings imply that excessive bandwidth allocation does not further enhance transmission efficiency and may lead to resource wastage, highlighting the need to consider other limiting factors. Although DMLA delivers the best transmission performance, it also shows the highest outage probability. This phenomenon arises because NOMA allows multiple UAVs to share a single subchannel, intensifying inter-group interference and reducing the per-subchannel transmission rate per UAV. Despite this, overall throughput

increases due to multi-subchannel multiplexing. In contrast, FTPC achieves the lowest outage probability, benefiting from its refined time-division strategy, which enables UAVs to dynamically adjust transmit power in response to environmental variations, thereby ensuring communication continuity.

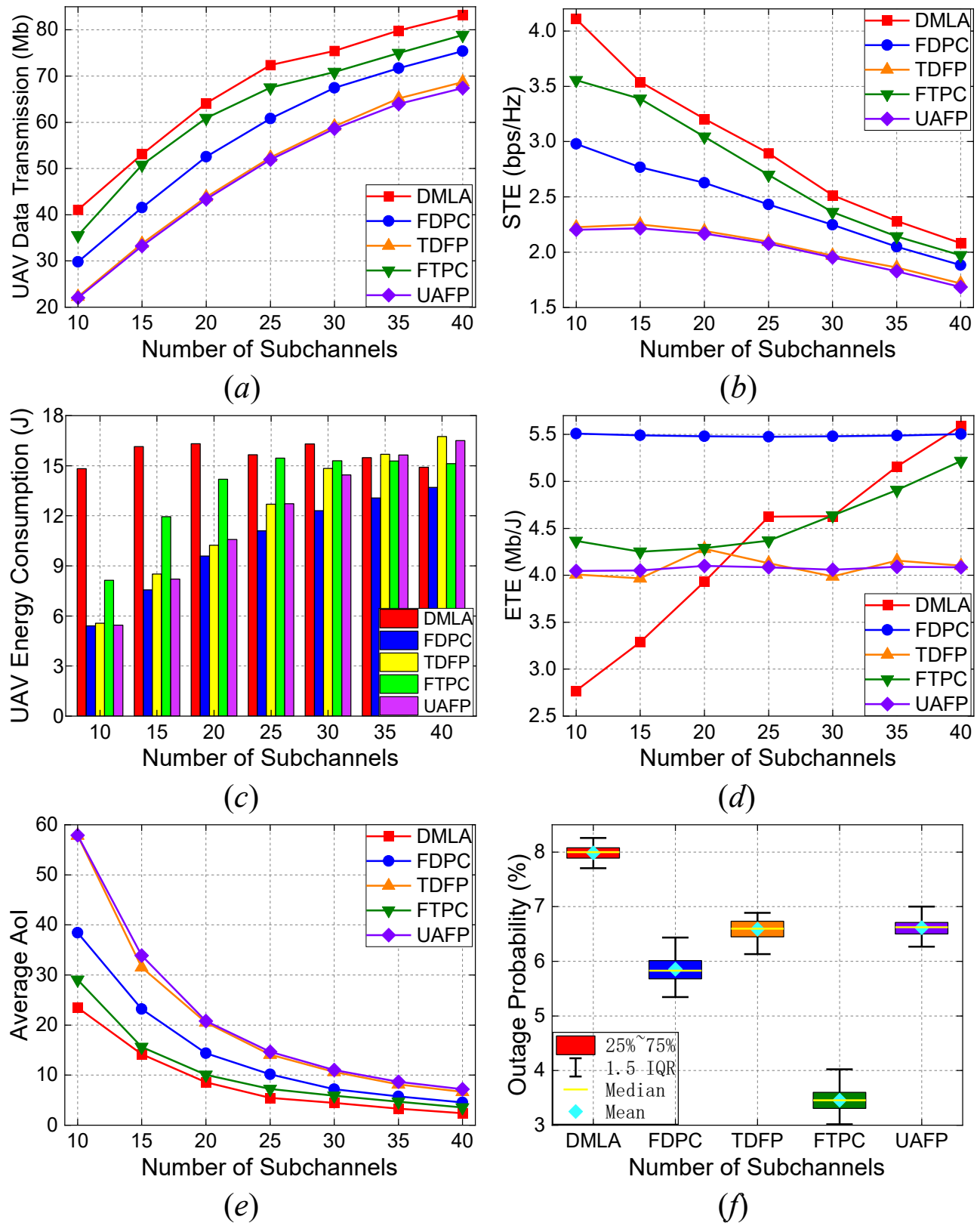

Fig. 5. Performance comparison of different schemes to $\mathbb{L}2$ under different number of subchannels.

*3) Resource Allocation Comparison of Different Integrated Solutions to* $\mathbb{GP}$

Fig.6(*a*), Fig.6(*b*), Fig.6(*c*), and Fig.6(*d*) illustrate the minimum number of SATs required for each scheme to achieve S-AoI proportions below 15%, 10%, 5%, and 1% under varying numbers of subchannels in a single-LEO scenario. First, as the number of subchannels increases, the transmission capability of each scheme improves, leading to a reduction in the AoI of data packets. This necessitates the deployment of additional SATs to decrease inter-SAT distance and consequently lower the S-AoI. Second, the required number of SATs is inversely related to transmission capability. IS-UAV-DMLA and IS-UAV-FTPC require more SATs than the other three schemes due to their superior transmission performance, which imposes stricter AoI requirements.

Finally, the figures indicate that as the S-AoI proportion decreases, the minimum number of required SATs increases. Notably, the increase in SATs needed to reduce the S-AoI proportion from 5% to 1% is greater than that required to reduce it from 15% to 10%. This suggests that during periods when UAVs are waiting for SAT services, their AoI continues to rise. Long waiting times mean that deploying a small number of additional SATs can significantly enhance transmission performance. Conversely, as waiting times shorten and UAVs gain more frequent data transmission opportunities, the marginal performance improvement gained from adding more SATs diminishes. In practical applications, communication scenarios with low AoI-efficiency requirements exhibit some tolerance for lower SAT coverage density. In contrast, AoI-sensitive real-time communication scenarios demand a denser SAT network to meet stringent performance criteria.

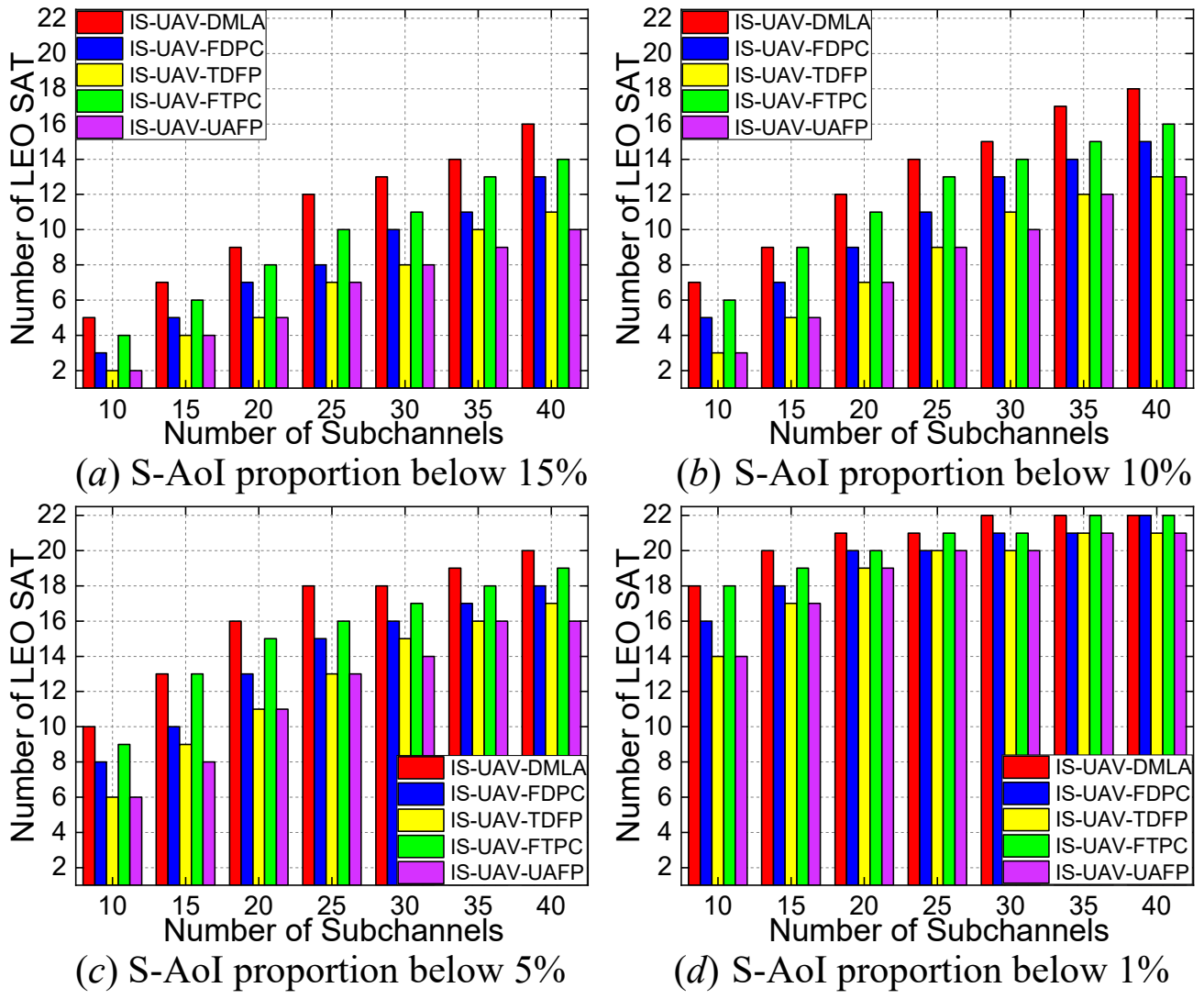

Fig. 6. The required numbers of SATs in a single-LEO under different S-AOI.

*4) AoI and Single-LEO SAT Density Expression*

In light of the aforementioned experimental results analysis, we attempt to derive the expected values of G2A-AoI, A2S-AoI, and S-AoI proportion, respectively.

*a) G2A-AoI Expected Value:* The probability that GT-$n$ is covered by UAV-$m$, the coverage set when UAV-$m$ covers GT-$n$, and the scheduling probability when GT-$n$ is covered by UAV-$m$ are denoted as $q_{n,m}^{C}$, $\widehat{\mathcal{N}}_m^G = \{n | c_{n,m} = 1, \forall n \in \mathcal{V}^G\}$, and $q_{n,m}^{S}$, respectively. $q_{n,m}^{S}$ is calculated as

$$q_{n,m}^{S} = \begin{cases} 1\,, & if\ \left|\widehat{\mathcal{N}}_m^G\right| \le Y^U, \\ \frac{\left|\{i | \varrho_i < \varrho_n, \forall i \in \widehat{\mathcal{N}}_m^G\}\right|}{\left|\widehat{\mathcal{N}}_m^G\right|}, & if\ \left|\widehat{\mathcal{N}}_m^G\right| > Y^U. \end{cases} \quad (32)$$

Given the possibility of multiple UAVs covering a single GT, directly calculating the GT scheduling probability is challenging. Therefore, we adopt the probability complement rule. First, we compute the probability that GT fails to connect to all UAVs, then subtract this value from 1 to obtain the successful scheduling probability. The scheduling interval of GT-$n$ follows a geometric distribution with an expectation of $\mathbb{E}[T_n^G] = \frac{1}{\left(1 - \prod_{m=1}^{M}(1 - q_{n,m}^{C} q_{n,m}^{S})\right)}$. Since AoI increases linearly between two successful transmissions and GT packets generation follows a Poisson distribution, the AoI of GT packets adheres to a discrete uniform distribution [48]. Consequently, the expected AoI value per packet for G2A communication can be expressed as

$$\mathbb{E}[A^G] = \frac{1}{2N} \sum_{n=1}^{N} min\left(\frac{\bar{d}_n^G}{\bar{D}_n^G}, 1\right)\left(\mathbb{E}[T_n^G] + 1\right), \quad (33)$$

where $\bar{d}_n^G$ is the average size of GT-$n$'s data packets, and $\bar{D}_n^G$ is the average data volume transmissible per time slot for GT-$n$ as calculated via (5). It can be inferred from (33) that coverage probability and transmission capability exert significant impacts on the expected value of G2A-AoI, which are primarily positively correlated with UAV path planning and transmission rate. Our IS-UAV architecture has demonstrated superior performance in this regard in the preceding comparison experiments. Additionally, we have optimized GT scheduling, which also affects G2A-AoI, by assigning each GT a priority $\varrho_n(t)$ that comprehensively balances AoI and fairness.

*b) A2S-AoI and Single-LEO SAT Density Expression:* According to single-LEO coverage model, the actual service duration of SAT- $l^k$ for the area of interest is $T^S_{m,l^k} = min(T^k, T_{m,l^k})$, and the waiting time of UAV-$m$ for SAT-$l^k$'s service is $T^W_{m,l^k} = max\left(T^k - T_{m,l^k}, 0\right)$, where $T^k = \frac{2\pi(d_E+h_k)}{L^S v^S}$ is the SAT interval of LEO-$k$. Therefore, based on the discrete uniform distribution, the expected AoI value per packet for A2S communication can be expressed as

$$\mathbb{E}[A^G] = \frac{1}{mL^S}\sum_{l^k\in\mathcal{V}^k}\sum_{m=1}^{M}\left(\frac{T^W_{m,l^k}}{T^k}\left(\frac{T^W_{m,l^k}+1}{2}+\frac{\bar{d}^U_m}{\bar{D}^U_m}\right)+\frac{T^S_{m,l^k}}{T^k}\frac{\bar{d}^U_m}{\bar{D}^U_m}\right), \tag{34}$$

where $\bar{d}^U_m$ is the average size of UAV-$m$ data packets, and $\bar{D}^U_m$ is the average data volume transmissible per time slot for UAV-$m$ as calculated via (16). Furthermore, the expected value of S-AoI proportion for data packets is computed as

$$\mathbb{E}[\partial] = \frac{\sum_{l^k\in\mathcal{V}^k}\sum_{m=1}^{m}\left(T^W_{m,l^k}\right)}{mL^S\mathbb{E}[A^G]+\mathbb{E}[A^G]}. \tag{35}$$

Let the S-AoI proportion range $\boldsymbol{\partial}^S$ be $[\partial_{min}, \partial_{max}]$, so that $\partial_{min} \leq \mathbb{E}[\partial] \leq \partial_{max}$ . This allows the calculation of the minimum and maximum expected values of $L^S$, based on which $L^{max}$ and $L^{min}$ in Step 5 of Algorithm 2 are set. Accordingly, Algorithm 2 can quickly find the number of single-LEO SATs matching the actual communication scenario within a short time, while avoiding algorithm search failures caused by $L^{max}$ and $L^{min}$ exceeding the feasible area.

## VII. CONCLUSIONS

To address communication disruptions caused by GBS failures in post-disaster areas, we proposed the PDPCIN framework, which integrates UAV-enabled WDC/WET and leverages LEO SATs to relay data to the nearest operational GBS. To ensure fundamental connectivity while collaboratively optimizing AoI, energy efficiency, and spectrum utilization, we designed three key components of PDPCIN: the AFTU mechanism for dynamic GT type updates, the IS-UAV architecture for simultaneous WDC and WET operations, and the DMLOA strategy for coordinated scheduling across multiple UAVs and LEO SATs.

Given the MINLP nature of the global problem, we developed the HHGNN framework, which models heterogeneous devices and their communication relationships as a hierarchical heterogeneous graph using the customized G3M. HHGNN decomposes the global problem into two layers: one layer ($\mathbb{L}1$), which optimizes UAV 3D trajectories and WET decisions; and the other layer ($\mathbb{L}2$), which focuses on UAV-LEO scheduling, power control, and subchannel allocation based on the outcomes from $\mathbb{L}1$. To further investigate how the number of SATs within a single LEO affects AoI and spectrum utilization under S-AoI constraints, we proposed the S-LSDO algorithm, which employs a binary search-based iterative optimization approach to determine the optimal LEO SATs count.

Extensive simulation results have demonstrated that the proposed approach outperforms existing benchmarks in jointly optimizing AoI, energy efficiency, and spectrum utilization. Based on this analysis, we further derived analytical expressions for the expected values of AoI and S-AoI proportion to guide resource allocation within PDPCIN. In future work, we will explore more realistic scenarios, such as signal degradation among GTs due to interference, and investigate the application of next-generation multiple access techniques to mitigate inter-UAV interference and enhance bandwidth utilization. Additionally, we plan to integrate large language model-enhanced reinforcement learning to improve the representational capacity of GNNs and enhance the generalization capability of the algorithms.

ACKNOWLEDGEMENT

This work was supported in part by the National Natural Science Foundation of China (62272484) and the High Performance Computing Center of Central South University.